\documentclass[fleqn,usenatbib]{mnras}

\usepackage{newtxtext,newtxmath}
\usepackage[T1]{fontenc}

\DeclareRobustCommand{\VAN}[3]{#2}
\let\VANthebibliography\thebibliography
\def\thebibliography{\DeclareRobustCommand{\VAN}[3]{##3}\VANthebibliography}

\usepackage{graphicx}
\usepackage{subcaption}
\usepackage{amsmath}
\usepackage{xcolor}
\usepackage{float}
\usepackage{orcidlink}

\newcommand{\Rg}{R_{\rm g}}
\newcommand{\vect}[1]{\boldsymbol{#1}}

\title[X-ray Comptonization in a tilted-disc Kerr black hole system II]{X-ray Comptonization in an Accreting Kerr Black Hole System with a Tilted Disc and Thermal Corona: A General-Relativistic Monte Carlo Simulation. II. Polarization Signatures}

\author[Mainak Chatterjee \& Dipankar Bhattacharya]{
Mainak Chatterjee,$^{1}$\orcidlink{0009-0005-1036-4523}\thanks{mainak.chatterjee\_phd23@ashoka.edu.in; mainak216@gmail.com}
Dipankar Bhattacharya$^{1}$\orcidlink{0000-0003-3352-3142}\thanks{dipankar.bhattacharya@ashoka.edu.in}
\\
$^{1}$Department of Physics, Ashoka University, Sonipat, Haryana 131029, India
}

\date{Accepted XXX. Received YYY; in original form ZZZ}
\pubyear{\the\year{}}

\begin{document}
\label{firstpage}
\pagerange{\pageref{firstpage}--\pageref{lastpage}}
\maketitle

\begin{abstract}
Relativistic precession of the inner accretion flow is widely invoked to explain low-frequency quasi-periodic oscillations (QPOs) in black hole X-ray binaries, and the Relativistic Precession Model (RPM) is expected to imprint a periodic polarimetric signature on the emergent X-rays. However, X-ray polarimetry with the Imaging X-ray Polarimetry Explorer (IXPE) has begun to reveal polarization properties that simple precession and Comptonization pictures struggle to reproduce. Two examples are Swift\,J1727.8$-$1613, whose upper limits on the QPO-phase polarization modulation lie well below the naive RPM prediction, and IGR\,J17091$-$3624, whose polarization degree of $\sim\!9\%$ in the hard state is well above the $\sim\!4\%$ seen in other hard-state black hole binaries observed by IXPE. Such anomalies call for a self-consistent framework in which the effects of geometry, scattering, and strong gravity on the polarization can be isolated and quantified. In Paper~I of this series we presented a general-relativistic Monte Carlo (GRMC) radiative transfer model of X-ray reprocessing in a misaligned, warped two-zone accretion disc around a Kerr black hole, surrounded by a hot optically thin corona, and used it to compute the resulting energy spectra, photon path-length and arrival-time distributions, and X-ray spectral lags. Here we extend the transport engine of Paper~I to track the full photon polarization state, adding a Walker--Penrose parallel-transport formalism and a polarized treatment of Compton and disc scattering. We compute the energy-resolved degree of polarization (DOP) and angle of polarization (AOP) of the composite system, and their modulation with precession phase, as functions of the coronal optical depth, outer-disc albedo, photon-injection radius, and observer inclination. We find that gravitational light bending raises the mean DOP and increases the polarization modulation expected from a precessing inner flow, whereas coronal scattering and reflection from the misaligned outer disc lower the DOP and suppress the modulation. The steady, comparatively bright polarization signal of the outer disc further dilutes the phase-dependent inner-disc modulation. The relative strength of these competing effects determines the observed polarization, so that the same model can produce both a low, nearly unmodulated signal and a high, strongly modulated one. As illustrations, a scattering-rich, reflection-dominated configuration yields a modulation broadly consistent with the non-detections in Swift\,J1727.8$-$1613, while a configuration with little secondary scattering and strong light bending yields a high DOP of the kind seen in IGR\,J17091$-$3624. These results provide a general framework for interpreting anomalous polarimetric behaviour in accreting black hole systems, and testable predictions for constraining black hole parameters, coronal properties, and disc-warp geometry with current and future X-ray polarimetry missions such as IXPE, XPoSat, and eXTP.
\end{abstract}


\section{Introduction}
\label{sec:intro}

Most black hole X-ray binaries (BHXRBs) show, during outburst, a narrow low-frequency quasi-periodic oscillation (QPO) in their hard and hard-intermediate spectral states \citep{BelloniMotta2016, IngramMotta2019}. The relativistic precession model (RPM) ties the strongest and most common of these, the type-C QPO, to Lense--Thirring precession \citep{LenseThirring1918} of a hot inner accretion flow that is misaligned with the black hole spin axis and precesses as a solid body \citep{StellaVietri1998, IngramDoneFragile2009}. Such misalignment naturally produces a warped, Bardeen--Petterson-type geometry \citep{BardeenPetterson1975}, in which the inner flow precesses while a cooler outer disc remains at its own, different tilt. In Paper~I of this series \citep[hereafter Paper~I]{ChatterjeeBhattacharya2026a} we presented a general-relativistic Monte Carlo (GRMC) radiative transfer model of this two-zone system, including the disc--corona geometry, the photon transport engine, and the observer pipeline, and used it to compute the resulting energy spectra, photon path-length and arrival-time distributions, and X-ray spectral lags as functions of the physical parameters of the system. Here we extend that model to compute its polarization signatures.

Timing and spectra alone have not been able to decide between the RPM and competing explanations for the type-C QPO, and polarimetry provides an independent constraint. Even for a static, unperturbed accretion disc, the X-ray continuum emerging from a Comptonizing corona is expected to be intrinsically polarized \citep{LightmanShapiro1975}. Radiation scattered in a corona or a disc atmosphere is polarized, and the observed degree (DOP) and angle (AOP) of polarization depend on the scattering geometry and on general-relativistic effects along the photon path, which rotate the polarization vector \citep{StarkConnors1977, ConnorsPiranStark1980}. If the QPO arises from precession, the changing orientation of the emitting region should modulate the DOP and AOP with QPO phase. \citet{Ingram2015} computed this by parallel-transporting polarization vectors through the Kerr metric for a precessing Comptonizing flow. For inclinations of $30^{\circ}$--$60^{\circ}$ the predicted modulation is at least $\sim\!0.5\%$ in DOP and $\sim\!4^{\circ}$ in AOP, as quoted by \citet{Zhao2024}.

The Imaging X-ray Polarimetry Explorer (IXPE), launched on 9 December 2021, measures polarization in the 2--8\,keV band \citep{Weisskopf2022}. Its first result on a BHXRB was for Cyg\,X$-$1 in the hard state, with DOP\,$=4.01\pm0.20\%$ and an AOP aligned with the radio jet \citep{Krawczynski2022}, consistent with a hot medium extended in the plane of the disc rather than a jet-base or lamp-post source. A similar DOP of about 4\% with a jet-aligned AOP was later measured in Swift\,J1727.8$-$1613 through the hard and hard-intermediate states \citep{Veledina2023, Ingram2024}.

Swift\,J1727.8$-$1613 was bright enough to test the precession prediction directly. \citet{Zhao2024} folded the IXPE data from the hard-intermediate state on a type-C QPO at 1.34\,Hz (fractional rms $\sim\!12\%$). The photon index varied strongly with QPO phase, but the DOP and AOP did not. Sine fits gave amplitudes of $0.2\pm0.2\%$ in DOP and $1.4^{+1.6}_{-1.0}$ degrees in AOP,consistent with no modulation and well below the expectation from the simple precession model. They note that a high inclination could weaken the modulation, since photons from the far side of the flow pass close to the black hole and are depolarized, but it would also push the mean DOP to $\gtrsim 7\%$, above the observed $\sim\!4\%$.

The polarization of hard-state BHXRBs observed by IXPE is, however, not uniformly low. \citet{Ewing2025} reported the first IXPE detection of the transient black hole low-mass X-ray binary IGR\,J17091$-$3624 in the corona-dominated hard state, with a DOP of $9.1\pm1.6\%$ ($5.2\sigma$) and an AOP of $83^{\circ}\pm5^{\circ}$ in the 2--8\,keV band. This DOP is much higher than the $\sim\!4\%$ seen in Cyg\,X$-$1 and Swift\,J1727.8$-$1613. In a Comptonization scenario, they argue that such a high DOP requires a very favourable coronal geometry, a high inclination (supported by a dip in the light curve), and possibly a mildly relativistic outflow and/or scattering in an optically thick wind. The polarization of hard-state BHXRBs therefore spans a wide range, from the $\sim\!4\%$ of the two sources above to $\gtrsim9\%$. These two sources illustrate that the observed polarization can depart from simple expectations in opposite directions: a modulation much weaker than the simple precession prediction in Swift\,J1727.8$-$1613, and an unusually high mean DOP in IGR\,J17091$-$3624. As IXPE observes more sources, further departures of this kind are likely to appear. Interpreting them calls for a self-consistent model in which the effects of geometry, scattering, and strong gravity on the polarization can be isolated and quantified.

Such departures may partly come from the modelling rather than from the physics of precession itself. The predictions above treat the precessing flow in isolation, with simplified assumptions for the Comptonization that produces the emitted spectrum \citep{Ingram2015}. In a real system the flow is surrounded by hot plasma that scatters the photons again on their way out, which can depolarize them. Part of the radiation reaches the outer disc and returns as reflection, and everything is seen through the strong-field geometry of the Kerr spacetime. Each of these can change both the mean DOP and its modulation, in either direction. An outer disc misaligned with the spin is also illuminated from a direction that changes over the precession cycle, so its reflected light should carry a phase-dependent polarization of its own. The geometry of the corona is also not settled: for instance, an outflowing rather than static corona has recently been invoked to explain the IXPE polarization signal from Cyg\,X$-$1 \citep{Poutanen2023}, underscoring that the assumed coronal geometry is itself an important, and adjustable, ingredient in these predictions.

Polarized radiative transfer around Kerr black holes has been modelled extensively. The basic physics of polarization from scattering atmospheres goes back to \citet{Chandrasekhar1960}, with Comptonizing slabs treated by \citet{SunyaevTitarchuk1985} and \citet{PoutanenSvensson1996}. For the disc emission, the ray-tracing calculations of \citet{Dovciak2004, Dovciak2008, Dovciak2011}, \citet{Li2009} and \citet{SchnittmanKrolik2009} include the Kerr-metric effects. For coronae, \citet{SchnittmanKrolik2010}, \citet{Beheshtipour2017}, \citet{ZhangDovciakBursa2019, Zhang2022} and \citet {KrawczynskiBeheshtipour2022} have explored a range of geometries.

Most directly relevant to the present work, \citet{Abarr2020} computed the energy spectra and polarization of a Bardeen--Petterson-type warped disc with a general-relativistic ray-tracing code that parallel-transports the polarization vector. Their disc consists of two rigid planes: an inner disc aligned with the black hole spin and an outer disc tilted by $15^{\circ}$ beyond a break radius of $8\,\Rg$ (for $a=0.9$ and $M_{\rm BH}=10\,\mathrm{M_\odot}$). The code includes single and multiple reflections of returning radiation off both disc segments, treated with the Chandrasekhar scattering formalism. They found that the polarization angle depends strongly on the azimuthal viewing angle of the misaligned system: it follows that of the outer disc at low energies, where direct outer-disc emission dominates, and carries the imprint of the rotation between the two disc axes at high energies, where reflected photons dominate. Given an independent measurement of the jet direction, the polarization angle can therefore place a lower limit on the disc misalignment. They also noted that polarimetry could test whether the precession of the inner disc drives the QPOs \citep{Ingram2015}. Their approach, however, considers only thermal disc emission and its reflection, for a static (non-precessing) warp and a fully reflecting disc, and it contains neither a Comptonizing corona nor a time-dependent inner flow. To our knowledge, none of the works discussed above treats, within a single framework, a tilted inner accretion flow in the Kerr spacetime, a surrounding corona, and reflection from a misaligned outer disc, while accounting for the changing viewing geometry associated with inner-flow precession in a single calculation, which is what a comparison with QPO-phase-resolved data would need.

In this paper we extend the GRMC model of Paper~I to close this gap. The photon transport engine, disc--corona geometry, and observer pipeline are exactly as described there. Here we add a polarization 4-vector to every photon, parallel-transported along its geodesic using the Walker--Penrose conserved quantity \citep{WalkerPenrose1970}, and track its evolution through polarized Compton scattering in the corona, following the analytic framework for polarized Comptonization in a spherical corona of \citet{Dove1997}, and through polarized reflection off the tilted outer disc. The outputs are the energy- and precession-phase-resolved DOP and AOP of the composite system.

We use this model to investigate how coronal scattering, gravitational light bending, and reflection from the misaligned outer disc shape both the mean DOP and its modulation with precession phase. In our approach, the tilted inner flow and the surrounding corona are treated as stationary configurations, and the effects of precession are modelled by varying the observer's viewing direction relative to the flow. Because these effects can raise or lower the polarization, the model is not tailored to a single source: it provides a general framework for interpreting anomalous polarimetric behaviour. We map how the DOP and AOP curves depend on the parameters of the system, so that both non-detections and high-DOP detections can be turned into constraints on them. We use Swift\,J1727.8$-$1613 and IGR\,J17091$-$3624 as illustrative cases, showing how the same ingredients can produce a near-null modulation and an unusually high DOP.

By combining the inner flow, corona, and misaligned outer disc developed in Paper~I with the polarization formalism developed here, this paper links timing, spectroscopy, and polarimetry through one set of parameters and gives predictions that can be checked against the IXPE data now in hand. Confirming or ruling out a signal at this level will eventually need simultaneous polarimetry, timing, and broadband spectral coverage \citep{Zhao2024}, which is exactly what the polarimetry-focusing arrays of the enhanced X-ray Timing and Polarimetry mission (eXTP; \citealt{Zhang2019,Zhang2025eXTP}), are designed to provide.

The paper is organized as follows. Section~\ref{sec:system_recap} recaps the system and the radiative transfer model of Paper~I. Section~\ref{sec:polarized_transfer} describes the polarized radiative transfer. Section~\ref{sec:results_polarization} presents the results: the fiducial model, the energy and inclination dependence of the polarization, the separate effects of scattering and light bending, and the polarization of the precessing disc and of the full system. Our conclusions are given in Section~\ref{sec:conclusions}.
\section{Recap: The System and Radiative Transfer Model}
\label{sec:system_recap}
\begin{figure}
    \centering
    \includegraphics[width=0.95\linewidth]{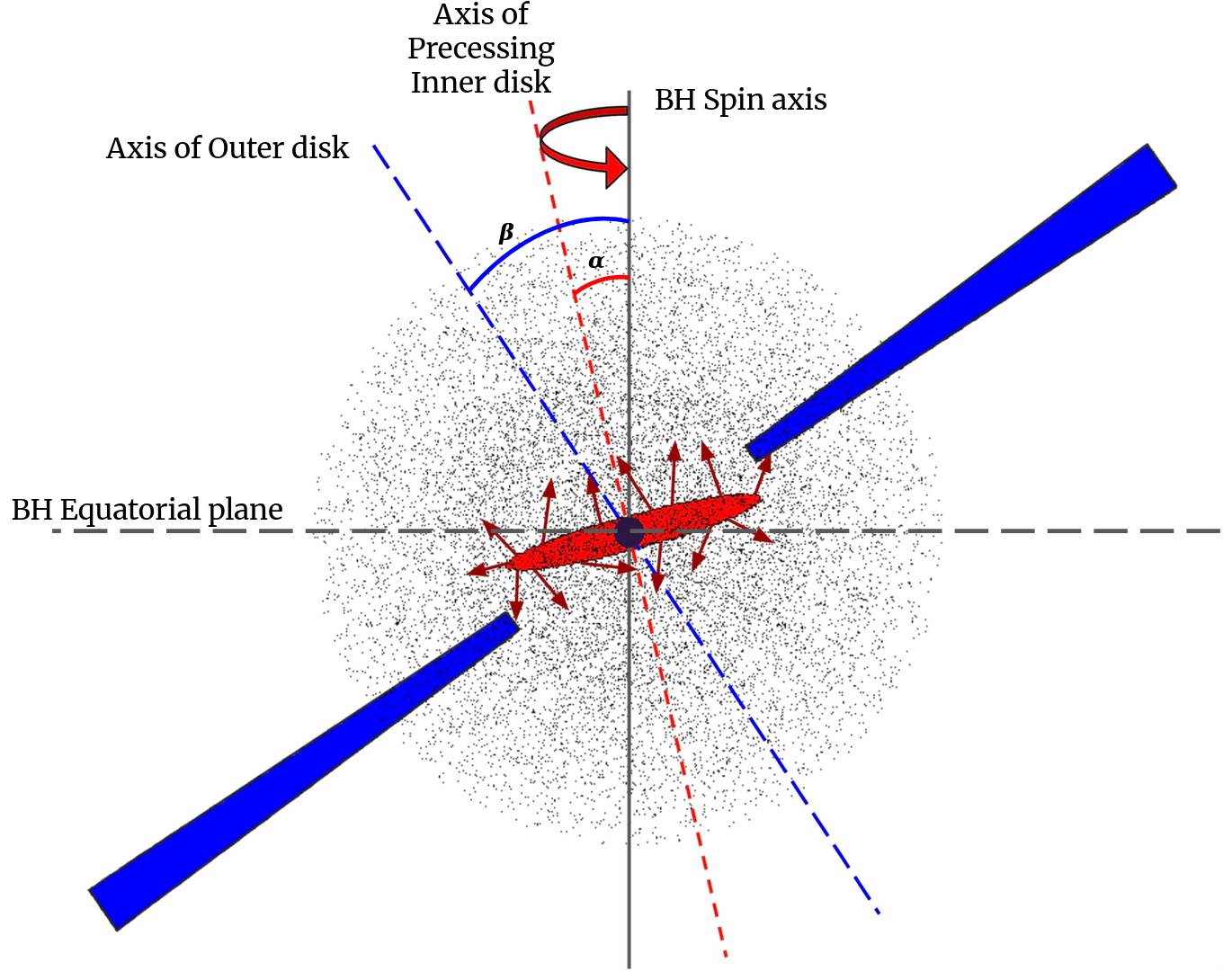}
    \caption{Schematic diagram of the proposed system. It comprises the BH--corona system, the precessing inner disc (red) and the stationary outer disc (blue), both tilted at different angles with respect to the BH spin axis. Here $\alpha$ and $\beta$ are $\theta_{\rm in}$ and $\theta_{\rm out}$ respectively.}
    \label{fig:system}
\end{figure}
The physical system (Figure \ref{fig:system}) is identical to that of Paper~I: a Kerr black hole of
mass $M_{\rm BH}$ and spin $a$; a hot inner accretion flow tilted by an angle
$\theta_{\rm in}$ from the black hole spin axis $\hat{z}_{\rm BH}$, surrounded by a
hot, spherically symmetric, optically thin Comptonizing corona; and a
stationary, cooler outer disc tilted by an angle $\theta_{\rm out}$, truncated at its own
Bardeen--Petterson warp radius $r_{\rm warp}$ and modelled analytically with the
Shakura--Sunyaev \citep{ShakuraSunyaev1973} thin-disc solution. Photons
are injected from the inner disc surface, propagate on null geodesics in
Boyer--Lindquist coordinates, Compton-scatter in the corona using a weighted
Monte Carlo method, and may be reflected by the outer disc before escaping to
a distant observer at inclination $i$ and azimuth $\phi_{\rm obs}$. The
derivation of the metric, geometry, geodesic integration, event detection, and
the flux and lag observer pipeline is given in Paper~I and is not repeated here.

What Paper~I did not track is the polarization state of the photon. Each photon
is additionally assigned, at emission, a polarization 4-vector $\epsilon^\mu$ that is
spacelike, unit-normed, and transverse to the photon 4-momentum $p^\mu$:
\begin{equation}
g_{\mu\nu}\epsilon^{\mu}\epsilon^{\nu} = 1, \qquad
g_{\mu\nu}p^{\mu}\epsilon^{\nu} = 0.
\end{equation}
The remainder of this paper describes how this vector is assigned,
transported, and scattered, and how it enters the observed Stokes spectrum.

\section{Polarized Radiative Transfer}
\label{sec:polarized_transfer}

\subsection{Initial polarization}
\label{sec:polarization_choice}

Photons are emitted from the inner disc along $\hat{n}_{\rm phot}$, with an initial energy drawn from a Planck distribution at the local disc temperature. Their polarization state follows the Chandrasekhar law for a semi-infinite electron-scattering atmosphere \citep{Chandrasekhar1960, ViironenPoutanen2004}, in which the electric vector of the emergent radiation is perpendicular to the plane containing the photon direction and the disc normal. The preferred polarization direction is therefore
\begin{equation}
\label{eq:f_perp}
  \hat{f}_{\perp}=\frac{\hat{n}_{\rm phot}\times\hat{n}_{\rm inner}}{\left|\hat{n}_{\rm phot}\times\hat{n}_{\rm inner}\right|},
\end{equation}
where $\hat{n}_{\rm inner}$ is the unit normal of the inner disc. This vector lies in the disc plane and is perpendicular to $\hat{n}_{\rm phot}$. The degree of polarization depends on the angle between $\hat{n}_{\rm phot}$ and $\hat{n}_{\rm inner}$ through
\begin{equation}
\label{eq:mu_DOP}
  \mu = |\hat{n}_{\rm phot}\cdot\hat{n}_{\rm inner}|, \qquad
  \mathrm{DOP} = \mathrm{DOP}_{\rm init}\,
  \frac{1-\mu}{1+3.582\,\mu},
\end{equation}
with $\mathrm{DOP}_{\rm init}=0.1171$ the maximum polarization, reached for edge-on escape ($\mu\to0$); the DOP vanishes for face-on escape. A uniform deviate $u$ then selects the polarization state of each photon: the photon is polarized along $\hat{f}_{\perp}$ if $u<\mathrm{DOP}$, and otherwise along a random direction in the plane perpendicular to $\hat{n}_{\rm phot}$. Random-branch photons average to zero net $Q$ and $U$, so the ensemble reproduces the desired DOP, while $\hat{\epsilon}\perp\hat{n}_{\rm phot}$ holds by construction. In the face-on limit, where the cross product in Eq.~\eqref{eq:f_perp} vanishes and the DOP is in any case close to zero, the random branch is used. The polarization vector is then passed to the photon transport, where it is converted to a Boyer--Lindquist 4-vector using the same ZAMO-tetrad construction as the photon momentum in Paper~I, followed by a Gram--Schmidt re-orthogonalization.

\subsection{Polarization transport in Kerr spacetime}
\label{sec:wp_reconstruction}

Along a null geodesic the polarization vector is parallel transported. Rather than integrating the transport equation, we use the Walker--Penrose constant \citep{WalkerPenrose1970}, a complex scalar $\kappa_{\rm WP}=K_1+iK_2$ built from the photon 4-momentum $k^\mu$ and the polarization vector $f^\mu$, which is conserved along the geodesic, $\mathrm{d}\kappa_{\rm WP}/\mathrm{d}\lambda=0$. We evaluate $\kappa_{\rm WP}$ at the emission point and, at each subsequent point where the polarization is needed, reconstruct $f^\mu$ from this conserved value together with the transversality condition $k_\mu f^\mu=0$ and a temporary gauge choice $f^t=0$. The resulting linear system for the spatial components is solved in closed form and the vector is renormalized. The explicit expression for $\kappa_{\rm WP}$ and the reconstruction follow the standard formulation of \citet{WalkerPenrose1970} and \citet{ConnorsPiranStark1980}. As a consistency check, $\kappa_{\rm WP}$ is recomputed from the reconstructed vector and compared with its conserved value, and the recovered $f^t$ is required to vanish.

\subsection{Polarized Compton scattering in the corona}
\label{sec:compton}

Whenever a geodesic step ends inside the corona, the polarization vector is carried through the same ZAMO-tetrad, Lorentz-boost, and scattering pipeline as the photon momentum in Paper~I, with the following additions.

\paragraph{Boost to the electron rest frame.}
Treating the photon as a plane wave with $\mathbf{B}=\hat{k}\times\boldsymbol{\epsilon}$, the polarization (electric-field) vector is boosted with the electromagnetic field tensor, giving in the electron rest frame (ERF)
\begin{equation}
\vect{\epsilon}'=\gamma_{e}\left(\vect{\epsilon}+\vect{\beta}_{e}\times\mathbf{B}\right)
-\frac{\gamma_{e}^{2}}{\gamma_{e}+1}\,\vect{\beta}_{e}\left(\vect{\beta}_{e}\cdot\vect{\epsilon}\right),
\end{equation}
which is then normalized to give $\hat{\epsilon}'_{\rm inc}$.

\paragraph{Scattering in the ERF.}
The outgoing direction is drawn from the polarization-dependent Thomson cross-section, $\mathrm{d}\sigma/\mathrm{d}\Omega'\propto1-(\hat{\epsilon}'_{\rm inc}\cdot\hat{n}'_{\rm scat})^2$, by rejection sampling of isotropic trial directions. This reduces to the unpolarized $1+\cos^2\Theta_{\rm scat}$ form of Paper~I when averaged over the incident polarization. The energy follows the Compton shift formula of Paper~I. The scattered polarization is the component of the incident vector perpendicular to the new direction,
\begin{equation}
\label{eq:thomson_polarization}
\hat{\epsilon}'_{\rm scat}\propto\hat{\epsilon}'_{\rm inc}-(\hat{\epsilon}'_{\rm inc}\cdot\hat{n}'_{\rm scat})\,\hat{n}'_{\rm scat},
\end{equation}
normalized to unit length, with transversality verified and enforced before the inverse boost.

\paragraph{Return to the lab frame.}
The inverse boost ($\vect{\beta}_e\to-\vect{\beta}_e$), applied with $\mathbf{B}'_{\rm scat}=\hat{n}'_{\rm scat}\times\hat{\epsilon}'_{\rm scat}$, returns the polarization to the flat-space Cartesian frame, from which the Boyer--Lindquist 4-vector is rebuilt via the inverse Jacobian and ZAMO tetrad as in Paper~I.

\subsection{Polarized reflection from the outer disc}
\label{sec:disc_polarization}

Reflection off the outer disc is modelled, as in Paper~I, as a single Compton scattering off electrons at rest, with the albedo $\rho_{\rm alb}$ applied as a weight. The outgoing direction is drawn from the polarization-dependent cross-section
\begin{equation}
\label{eq:thomson_dcs}
  \frac{\mathrm{d}\sigma}{\mathrm{d}\Omega}\propto1-\bigl(\hat{\epsilon}\cdot\hat{k}_{\rm sc}\bigr)^{2},
\end{equation}
by rejection sampling, the energy follows the Compton formula of Paper~I, and the scattered polarization is obtained by projecting the incident vector onto the plane perpendicular to $\hat{k}_{\rm sc}$ \citep{Chandrasekhar1960}, as in Eq.~\eqref{eq:thomson_polarization}. Transversality is checked to a tolerance of $10^{-10}$. As in Paper~I, this module deliberately omits the electron-velocity boost: the disc electrons are cooler and denser than those in the corona, so the dominant effect is angular redistribution and polarization rotation, and the albedo already parametrises the overall scattering efficiency. Our use of the Chandrasekhar projection for disc reflection is in the spirit of \citet{Abarr2020}, who scattered photons off a warped disc using the full Chandrasekhar reflection law for a semi-infinite electron-scattering atmosphere. There are two differences. First, our treatment is simpler, being a single Thomson scattering with projection of the polarization vector (Eq.~\ref{eq:thomson_polarization}) rather than the full Stokes reflection matrix. Second, \citet{Abarr2020} assumed that the disc reflects 100\% of the incident photons, whereas here the reflected fraction is controlled by the albedo $\rho_{\rm alb}$, which is varied as a free parameter in Section~\ref{sec:parameter_dependence_inclination} and Section~\ref{sec:precession_parameter_dependence}.

\subsection{Polarization of the outer disc's direct emission}
\label{sec:outer_disc_pol}

The outer disc's direct emission (Component~C of Paper~I) is not Monte Carlo sampled, but carries an intrinsic polarization computed analytically. For a semi-infinite plane-parallel electron-scattering atmosphere the emergent radiation is polarized perpendicular to the plane containing the disc normal and the line of sight \citep{Chandrasekhar1960}, with degree
\begin{equation}
\label{eq:outer_dop}
  \mathrm{DOP}_{\rm out} = P_{\rm max}
    \frac{1 - \mu_{\rm out}}{1 + 3.582\,\mu_{\rm out}},
\end{equation}
where $\mu_{\rm out}=|\hat{n}_{\rm out}\cdot\hat{n}_{\rm obs}|$ and $P_{\rm max}\approx11.71\%$ \citep{ViironenPoutanen2004}. The polarization angle in the disc's own frame is $\mathrm{AOP}_{\rm out}^{\rm disc}=90^{\circ}$, measured from the sky projection of $\hat{n}_{\rm out}$. Both quantities are energy-independent over the outer-disc band, since the Thomson cross-section is effectively constant, and carry no Monte Carlo uncertainty.

\subsection{Stokes parameters and the observer pipeline}
\label{sec:observer_polarization}

Escaping photons are sorted into Component~A (not reflected: unscattered and corona-scattered photons) and Component~B (reflected from the outer disc), and accepted through the shared patch observer of Paper~I. The patch is centred on the observer direction at polar angle $\theta_{\rm obs}$ from the spin axis $\hat{n}_{\rm BH}$ and extends $\pm\Delta\theta$ in polar angle and $\pm\Delta\phi$ in azimuth, so that its solid angle is $\Omega_{\rm patch}=2\Delta\phi\,[\cos(\theta_{\rm obs}-\Delta\theta)-\cos(\theta_{\rm obs}+\Delta\theta)]$. For each accepted photon with escape direction $\hat{k}$, a sky basis is built from the spin axis projected onto the plane perpendicular to $\hat{k}$,
\begin{equation}
\hat{e}_1=\frac{\hat{n}_{\rm BH}-(\hat{n}_{\rm BH}\cdot\hat{k})\,\hat{k}}{\left|\hat{n}_{\rm BH}-(\hat{n}_{\rm BH}\cdot\hat{k})\,\hat{k}\right|},\qquad
\hat{e}_2=\hat{k}\times\hat{e}_1 .
\end{equation}
The AOP is therefore measured counter-clockwise, as seen by the observer, from the projected spin axis. Anchoring the basis to $\hat{n}_{\rm BH}$, rather than to the precessing inner-disc normal, fixes the $\mathrm{AOP}=0$ convention at every precession phase, so any AOP swing reflects physics rather than a drifting reference frame. The polarization vector is used as stored, without re-projection or renormalization. For a photon with $\epsilon_1=\hat{\epsilon}\cdot\hat{e}_1$ and $\epsilon_2=\hat{\epsilon}\cdot\hat{e}_2$, the Stokes contributions are
\begin{equation}
s_0=\epsilon_1^2+\epsilon_2^2,\quad s_1=\epsilon_1^2-\epsilon_2^2,\quad s_2=2\epsilon_1\epsilon_2 .
\end{equation}

Each photon $k$ that escapes in energy bin $i$ and inside the patch contributes with the weight $W_k=w_k\,E_{{\rm esc},k}$, where $w_k$ is its photon weight (a photon number) and $E_{{\rm esc},k}$ its escape energy. Since $w_k\,E_{{\rm esc},k}$ is the energy carried by the photon, all flux-like sums are energy-weighted. For each component $c\in\{A,B\}$, the energy-weighted flux and Stokes sums in bin $i$ are
\begin{equation}
\label{eq:stokes_sums}
\begin{split}
\noindent
&I_{c,i}=\frac{1}{\Omega_{\rm patch}}\sum_{k\in c,\,i} w_k\,E_{{\rm esc},k},\\
&S_{j,c,i}=\frac{1}{\Omega_{\rm patch}}\sum_{k\in c,\,i} w_k\,E_{{\rm esc},k}\,s_{j,k}\quad(j=0,1,2),
\end{split}
\end{equation}
where $s_{j,k}$ are the per-photon Stokes contributions defined above, and the sums run over the photons of component $c$ that escape in bin $i$ and fall inside the observer patch. The common factor $1/\Omega_{\rm patch}$ converts the sums from energy per patch to energy per unit solid angle, so that results for observer windows of different size can be compared. Because $s_{0,k}=1$ for a unit-normed transverse polarization vector, $S_{0,c,i}=I_{c,i}$ up to numerical error. Since $\Omega_A=\Omega_B=\Omega_{\rm patch}$, the components add directly,
\begin{equation}
S_{j,{\rm inner},i}=S_{j,A,i}+S_{j,B,i}\quad(j=0,1,2),\qquad I_{{\rm inner},i}=I_{A,i}+I_{B,i}.
\end{equation}
The inner-disc polarization in each energy bin is
\begin{equation}
\begin{split}
\mathrm{DOP}_{\rm inner}(E_i)&=\frac{\sqrt{S_{1,{\rm inner},i}^{2}+S_{2,{\rm inner},i}^{2}}}{S_{0,{\rm inner},i}},\\
\mathrm{AOP}_{\rm inner}(E_i)&=\tfrac{1}{2}\operatorname{tan^{-1}}\!\left(S_{2,{\rm inner},i},\,S_{1,{\rm inner},i}\right),
\end{split}
\end{equation}
with the AOP in the range $(-90^{\circ},90^{\circ}]$. At this stage the sums are unscaled quantities per energy bin, and no energy-bin width enters the DOP and AOP, since they are ratios. The broadband DOP and AOP are formed from the same sums added over the energy bins, so they are likewise independent of the bin widths.

The statistical uncertainty of the flux follows from the same sums as
\begin{equation}
\sigma_{I,c,i}=\frac{1}{\Omega_{\rm patch}}\sqrt{\sum_{k\in c,\,i}\left(w_k\,E_{{\rm esc},k}\right)^2},
\end{equation}
and the uncertainties on the DOP and AOP are estimated from the effective sample size,
\begin{equation}
\sigma_{\rm DOP}\simeq\frac{1-\mathrm{DOP}^2}{\sqrt{N_{\rm eff}}},\qquad
\sigma_{\rm AOP}\simeq\frac{1}{2\sqrt{N_{\rm eff}}\,\mathrm{DOP}}\ \ (\mathrm{rad}),
\end{equation}
where $N_{\rm eff}=(\sum w)^2/\sum w^2$ is the Kish effective sample size \citep{Kish1965}, evaluated in each energy bin from the merged A and B photon sums. These are approximate expressions that are conservative (overestimate the error by up to a factor of $\sqrt{2}$) for $\mathrm{DOP}\lesssim10\%$. They also assume $\mathrm{DOP}\sqrt{N_{\rm eff}}\gg1$, below which the DOP is biased upwards.

\subsection{Combining inner- and outer-disc polarization}
\label{sec:component_C_frame}
\label{sec:combine_final}

The polarization angle of Component~C is conventionally defined relative to the sky projection of the outer-disc normal, $\hat{u}_{\rm out}$, whereas Components~A and~B use the $\hat{n}_{\rm BH}$-referenced frame, with sky projection $\hat{u}_{\rm BH}$. Component~C is therefore rotated into the shared frame by the angle $\chi$ that carries $\hat{u}_{\rm out}$ onto $\hat{u}_{\rm BH}$ about the line of sight,
\begin{equation}
\chi=\mathrm{tan^{-1}}\!\left[\hat{n}_{\rm obs}\cdot(\hat{u}_{\rm out}\times\hat{u}_{\rm BH}),\ \hat{u}_{\rm out}\cdot\hat{u}_{\rm BH}\right],
\end{equation}
giving
\begin{equation}
\label{eq:component_C_rotate}
\begin{split}
&\mathrm{AOP}_{\rm out}^{\rm sky}=\mathrm{AOP}_{\rm out}^{\rm disc}-\chi,\\
&S_{1,i}^{\rm out}=F^{\rm out}_{E,i}\,\mathrm{DOP}_{\rm out}\cos(2\,\mathrm{AOP}_{\rm out}^{\rm sky}),\\
&S_{2,i}^{\rm out}=F^{\rm out}_{E,i}\,\mathrm{DOP}_{\rm out}\sin(2\,\mathrm{AOP}_{\rm out}^{\rm sky}),
\end{split}
\end{equation}
The negative sign arises because $\chi$ rotates $\hat{u}_{\rm out}$ onto $\hat{u}_{\rm BH}$: a direction at angle $\mathrm{AOP}_{\rm out}^{\rm disc}$ from $\hat{u}_{\rm out}$ lies at $\mathrm{AOP}_{\rm out}^{\rm disc}-\chi$ from $\hat{u}_{\rm BH}$. Here $F^{\rm out}_{E,i}$ is the spectral flux density of the outer disc, in $\mathrm{W\,m^{-2}\,keV^{-1}}$, averaged over energy bin $i$. The rotation degenerates harmlessly if either axis is aligned with the line of sight.

The inner-disc sums are converted into spectral flux densities only at this point. Using the scaling factor \texttt{mc\_scale} of Paper~I and dividing by the width $\Delta E_i$ of each energy bin, the composite Stokes vector in energy bin $i$ is
\begin{equation}
\begin{split}
&I_{{\rm tot},i}=\frac{\texttt{mc\_scale}\,I_{{\rm inner},i}}{\Delta E_i}+F^{\rm out}_{E,i},\\
&S_{j,{\rm tot},i}=\frac{\texttt{mc\_scale}\,S_{j,{\rm inner},i}}{\Delta E_i}+S^{\rm out}_{j,i}\quad(j=1,2),
\end{split}
\end{equation}
in units of $\mathrm{W\,m^{-2}\,keV^{-1}}$. The division by $\Delta E_i$, together with \texttt{mc\_scale}, puts the inner-disc sums on the same footing (a spectral density in physical units) as the band-averaged outer-disc emission, and the statistical uncertainty of the inner-disc flux is converted in the same way. The composite polarization is
\begin{equation}
\mathrm{DOP}_{\rm tot}=\frac{\sqrt{S_{1,\rm tot}^2+S_{2,\rm tot}^2}}{I_{\rm tot}},\qquad
\mathrm{AOP}_{\rm tot}=\tfrac12\operatorname{tan^{-1}}(S_{2,\rm tot},S_{1,\rm tot}).
\end{equation}
Only the Monte Carlo component carries statistical uncertainty. The uncertainties on $S_{1,{\rm inner}}$ and $S_{2,{\rm inner}}$ are obtained from those on the flux, DOP and AOP of the inner disc, assumed uncorrelated, and are then propagated to $\mathrm{DOP}_{\rm tot}$ and $\mathrm{AOP}_{\rm tot}$ with the standard partial-derivative formulae. The outer disc enters only through the totals.
\section{Results: X-ray Polarization}
\label{sec:results_polarization}

Based on the spectral and timing results of Paper~I, we now investigate how the physical parameters of the model influence the polarization properties of the observed radiation. In particular, we focus on the energy dependence of the DOP and AOP, and examine how these quantities are modified by the coronal properties, outer-disc properties, and viewing geometry.

We first define the fiducial model (Section~\ref{sec:fiducial}). We then consider a fixed observer and examine the dependence of the DOP and AOP on photon energy, which provides a baseline polarization signature against which the effects of the different physical parameters can be compared. Next, we study the dependence on observer inclination, and then isolate two competing physical effects: the scattering multiplicity set by the coronal optical depth, and the effect of gravitational light bending in the absence of scattering. We then examine how the inclination-dependent signatures change with the reflection fraction and the photon-emission radius. Finally, we extend the analysis to a precessing inner accretion disc and study both the resulting polarization signatures and their dependence on the physical parameters.

\subsection{Fiducial model and parameter variations}
\label{sec:fiducial}

In this section we analyse the data generated by the simulation and present the main quantitative results. All simulations are built around a common fiducial model, whose parameters are described below. The system consists of a $10\,\mathrm{M_\odot}$ black hole with spin $a=0.9$, surrounded by a spherical corona of radius $R_{\rm c}=50\,\Rg$. The electron-density profile of the corona,
\begin{equation}
\label{eq:elec_density}
n_{\rm e}(r)=n_{{\rm e},0}\exp\!\left(-r/R_{\rm c}\right),
\end{equation}
is normalized to $n_{{\rm e},0}=1\times10^{22}\,\mathrm{m^{-3}}$.

The inner and outer discs are tilted away from the BH spin axis (the $Z$ axis) by $\theta_{\rm in}=10^{\circ}$ and $\theta_{\rm out}=40^{\circ}$ respectively, with both tilts lying in the $Y$--$Z$ plane. The outer disc spans the radial range $r_{\rm warp}=10\,\Rg$ to $r_{\rm out}=400\,\Rg$ and reflects incident photons with an albedo of $\rho_{\rm alb}=0.2$. Seed photons are emitted from an annular ring of radius $R_0=6.5\,\Rg$ and width $\delta R_0=0.1\,\Rg$ on the inner disc, and each simulation is launched with $N_{\rm phot}=1\times10^{7}$ photons from this region. The Shakura--Sunyaev temperature normalization is set by the accretion rate $\dot M = \dot{m} L_{\rm Edd}/(\eta c^2)$, for which we adopt a radiative efficiency $\eta=0.1$ and an Eddington ratio $\dot{m} \equiv L/L_{\rm Edd}=0.1$, broadly representative of a BHXRB in a bright accretion state.

The black hole mass and spin of our fiducial model match those used by \citet{Abarr2020} for their warped disc ($M_{\rm BH}=10\,\mathrm{M_\odot}$, $a=0.9$), which provides a convenient point of comparison. The disc configurations, however, are different. In their model the inner disc is aligned with the spin axis, the break radius is $8\,\Rg$, the outer disc is tilted by $15^{\circ}$, and the reflection is 100\%. In ours, the inner and outer discs are tilted by $\theta_{\rm in}=10^{\circ}$ and $\theta_{\rm out}=40^{\circ}$ respectively, the warp radius is $r_{\rm warp}=10\,\Rg$, and the reflection is set by the albedo $\rho_{\rm alb}=0.2$.

These fiducial values are the starting point for all runs. The observer is placed at an arbitrary reference distance of $d = 3.086\times10^{25}\,\mathrm{m}$ ($\simeq 1\,\mathrm{Gpc}$) from the source. To show how the observables depend on the physical parameters of the system, we vary one parameter at a time and keep all others at their fiducial values, unless stated otherwise. The varied parameter and its range, together with the observer position and window ($\Delta\theta$, $\Delta\phi$), are given in the caption of each figure.

The observer window extends $\pm\Delta\theta$ in polar angle and $\pm\Delta\phi$ in azimuth about the observer direction, and is quoted as $\Delta\theta\times\Delta\phi$. A larger window collects more photons and so reduces the Monte Carlo noise, but at the cost of angular resolution. Unless stated otherwise, the DOP and AOP shown in the figures of Sections~\ref{sec:inclination_baseline}--\ref{sec:parameter_dependence_inclination} and in the precession sections refer only to the Monte Carlo (inner-disc) component, without the analytic outer-disc emission.

\paragraph{Fitting functions.}
The DOP as a function of observer inclination $\theta$ is fitted with a form motivated by the Chandrasekhar law (Eq.~\ref{eq:mu_DOP}),
\begin{equation}
\label{eq:fit_incl}
\mathrm{DOP}(\theta)=A\,\frac{1-\mu}{1+3.582\,\mu}+C,\qquad \mu=|\cos(\theta-B)|,
\end{equation}
where $A$ is an amplitude, the offset $B$ accounts for the tilt of the effective emitting surface relative to the spin axis, and $C$ is a constant offset. Unless an uncertainty is quoted, $B$ was held fixed at $\pm\theta_{\rm in}=\pm10^{\circ}$: its sign depends on the observer azimuth, with $B=+10^{\circ}$ for $\phi_{\rm obs}=270^{\circ}$ (peak DOP at $\theta=100^{\circ}$) and $B=-10^{\circ}$ for $\phi_{\rm obs}=90^{\circ}$ (peak DOP at $\theta=80^{\circ}$). In Figures~\ref{fig:changing_inclination_effects} and \ref{fig:inclination_albedo_dependence} it was left free, and the fitted values are consistent with $\theta_{\rm in}$. The DOP as a function of observer (precession) phase $\phi_{\rm obs}$ is fitted with a sinusoid,
\begin{equation}
\label{eq:fit_prec}
\mathrm{DOP}(\phi_{\rm obs})=A\sin\!\left(\omega\,\phi_{\rm obs}+\varphi\right)+C,
\end{equation}
with amplitude $A$, angular frequency $\omega$, phase offset $\varphi$, and constant offset $C$.
The DOP, AOP and flux shown in the inclination and precession scans are broadband quantities: the Stokes parameters and fluxes of Section~\ref{sec:observer_polarization} are summed over the energy bins, weighted by $\Delta E_i$, over the range $10^{-2}$--$10^{3}$\,keV, and the DOP and AOP are then formed from these sums, with uncertainties added in quadrature over the bins. The plotted flux is the band flux $\sum_i F_{E,i}\Delta E_i$ in $\mathrm{W\,m^{-2}}$.
\subsection{Energy dependence of the DOP and AOP}
\label{sec:energy_dependence}

\begin{figure*}
\centering
\begin{subfigure}{0.32\textwidth}
    \centering
    \includegraphics[width=\linewidth]{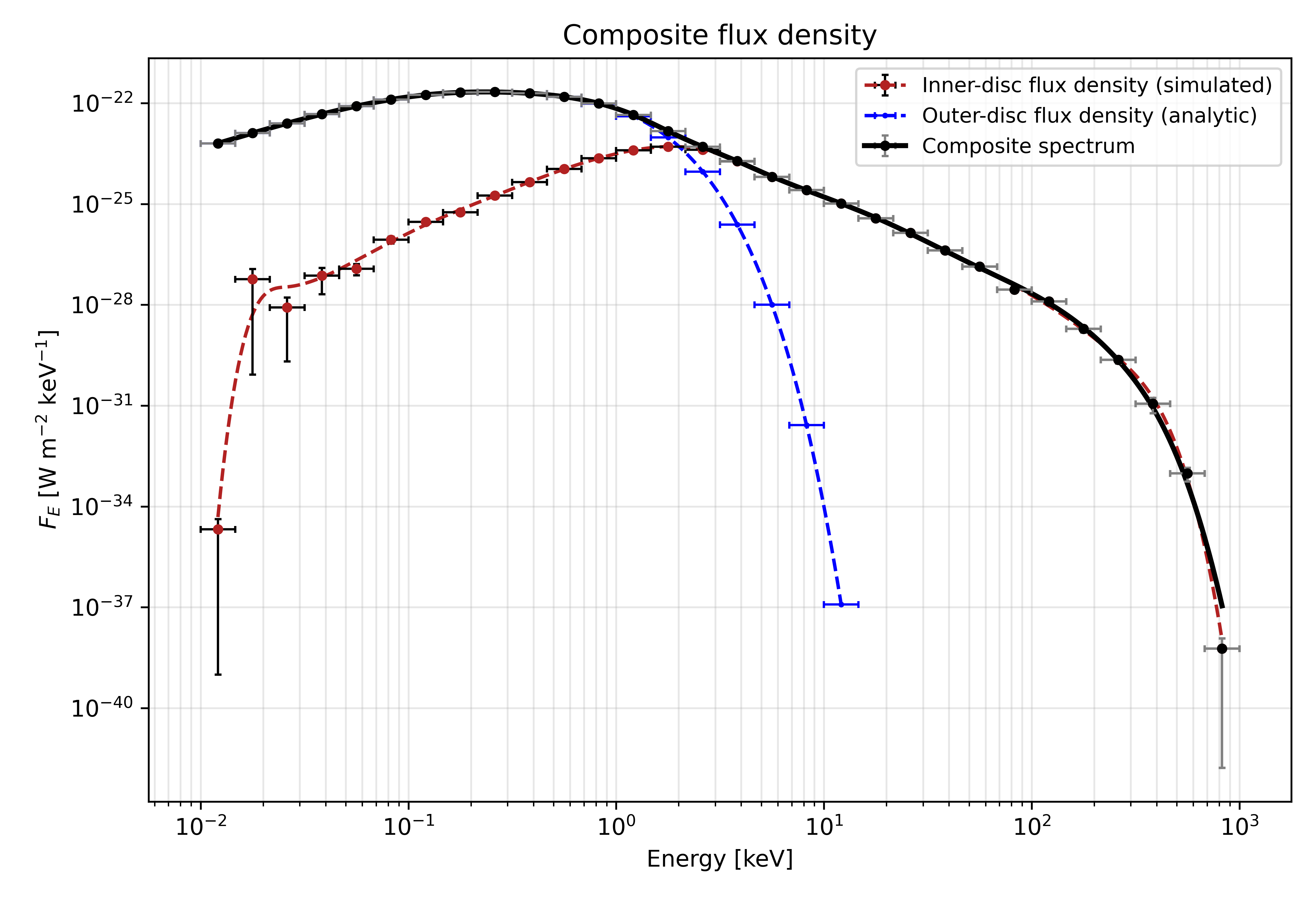}
    \caption{Composite flux density spectrum.}
    \label{fig:Composite_flux_density_spectrum}
\end{subfigure}
\hfill
\begin{subfigure}{0.32\textwidth}
    \centering
    \includegraphics[width=\linewidth]{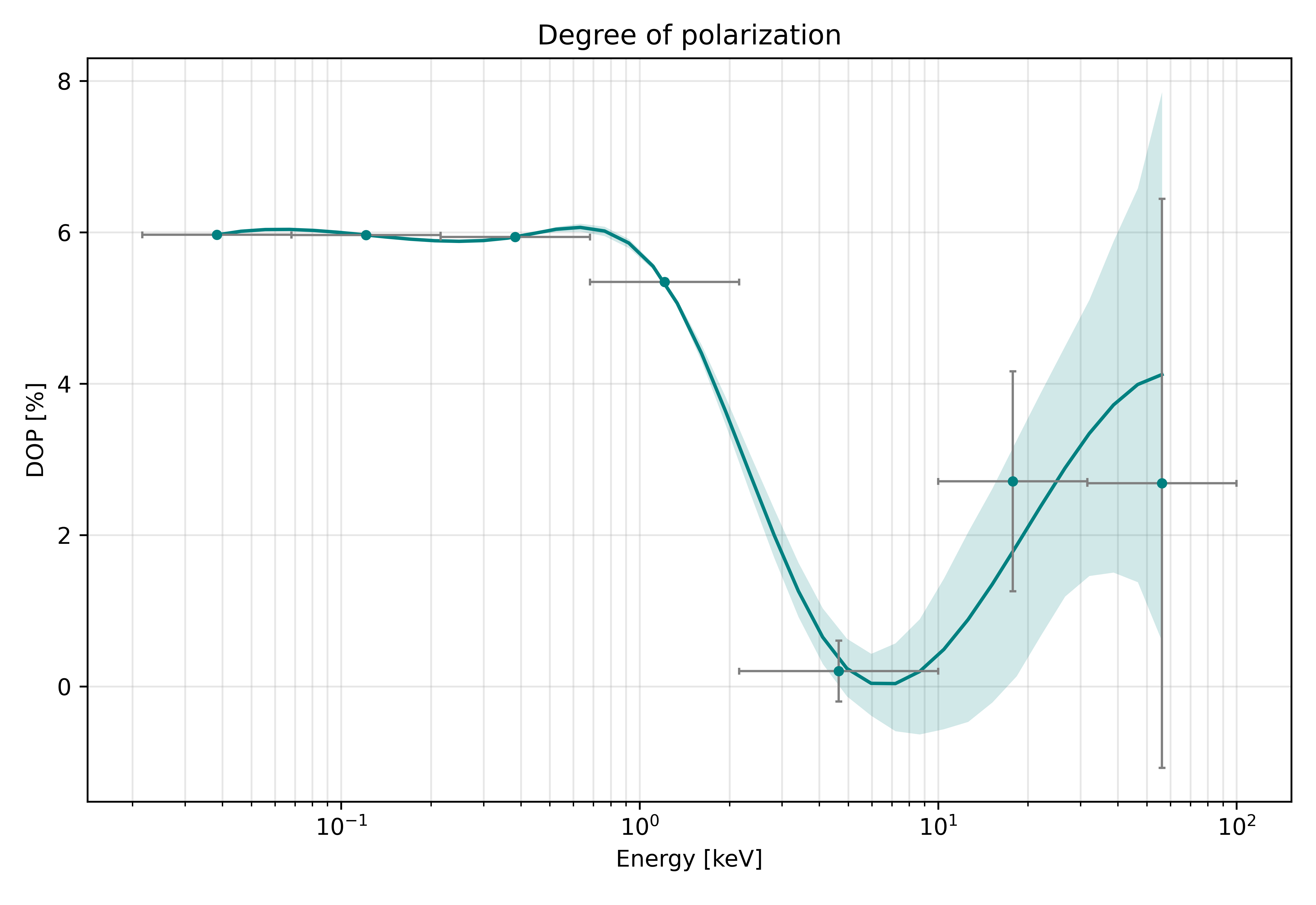}
    \caption{DOP as a function of photon energy.}
    \label{fig:DOP_E}
\end{subfigure}
\hfill
\begin{subfigure}{0.32\textwidth}
    \centering
    \includegraphics[width=\linewidth]{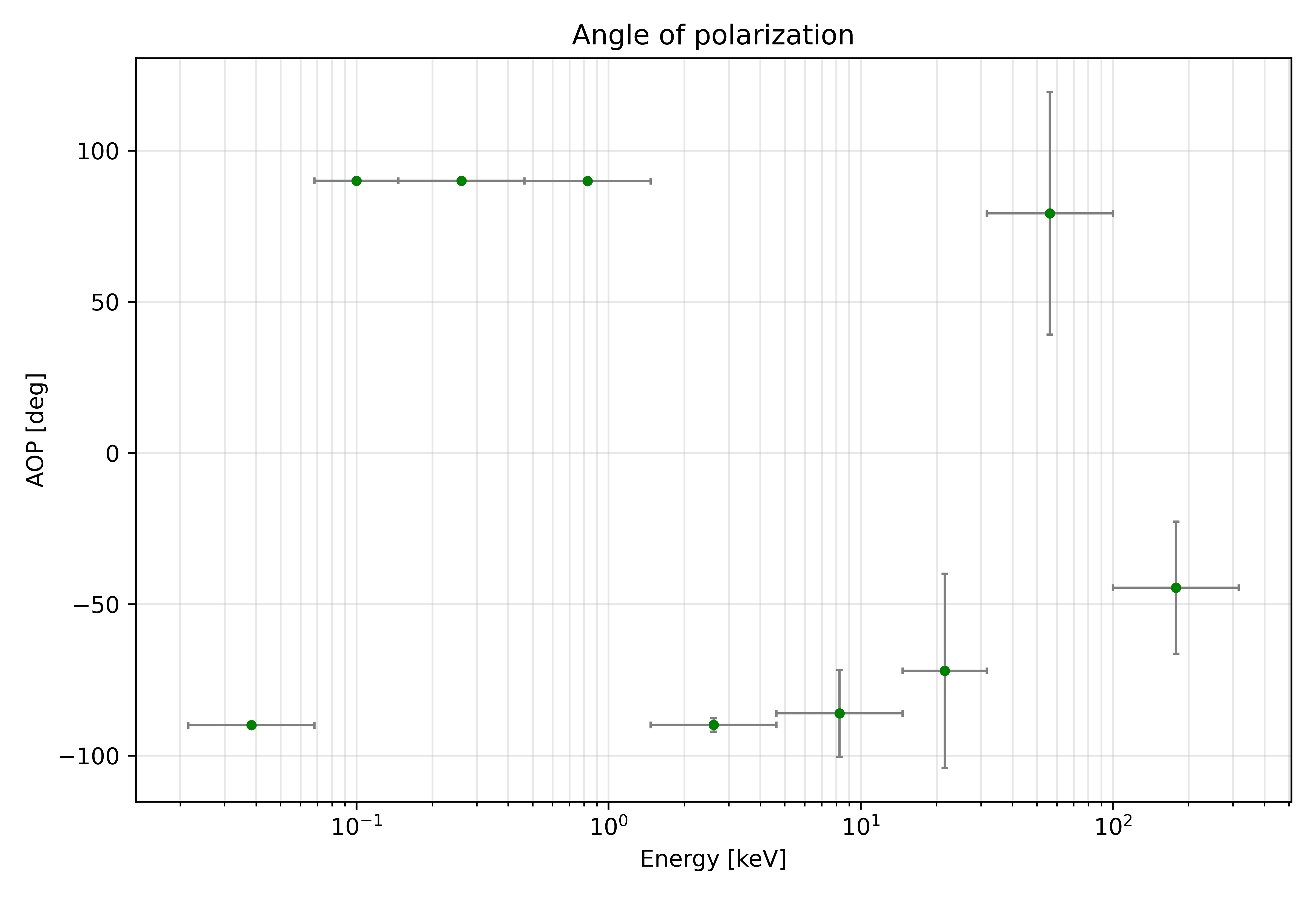}
    \caption{AOP as a function of photon energy.}
    \label{fig:AOP_E}
\end{subfigure}
\caption{Energy dependence of the flux density, DOP, and AOP of the composite (inner plus outer disc) system for the fiducial model. The observer is at an inclination of $40^{\circ}$ with respect to the BH spin axis and an azimuthal angle of $90^{\circ}$. The observer window is $\Delta\theta\times\Delta\phi=\pm3^\circ\times\pm3^\circ$ for the flux (a) and $\pm5^\circ\times\pm5^\circ$ for the DOP and AOP (b, c).}
\label{fig:polarization_energy}
\end{figure*}

We consider an observer fixed at an inclination of $40^{\circ}$ and an azimuthal angle of $90^{\circ}$, and investigate the energy dependence of the DOP and AOP. Figure~\ref{fig:polarization_energy} shows the resulting DOP and AOP across the photon-energy range. In Figure~\ref{fig:DOP_E}, we observe that at lower energies, where the outer disc dominates (as can be seen in Figure~\ref{fig:Composite_flux_density_spectrum}), the DOP is approximately 6\%. This is consistent with the expectation for this observer from Equation~\ref{eq:outer_dop}, given that the outer disc is tilted by $40^{\circ}$ relative to the BH spin axis (corresponding to an observer inclination of $80^{\circ}$ with respect to the outer disc) and that the flux from the outer disc undergoes no light bending.

As energy increases and the inner-disc photons, which undergo both scattering and light bending, begin to dominate, the DOP drops to approximately 0.5\% before rising again at higher energies. This is a deviation from the value of around 1.2\% expected from Equation~\ref{eq:mu_DOP}, given that the inner disc is tilted by $10^{\circ}$ relative to the BH spin axis (corresponding to an observer inclination of $\approx\!50^{\circ}$ with respect to the inner disc).
This provides the first evidence that scattering, together with light bending, affects the degree of polarization of the photons. We explore these effects in detail in the following sections.

Figure~\ref{fig:AOP_E} similarly shows that the AOP of the composite system remains at $90^{\circ}$, as expected, when the outer disc dominates. However, once the photons from the inner disc begins to dominate the spectrum, the AOP tends to bend inward.

\subsection{Dependence on observer inclination}
\label{sec:inclination_baseline}

\begin{figure*}
\centering
\begin{subfigure}{0.33\textwidth}
    \centering
    \includegraphics[width=\linewidth]{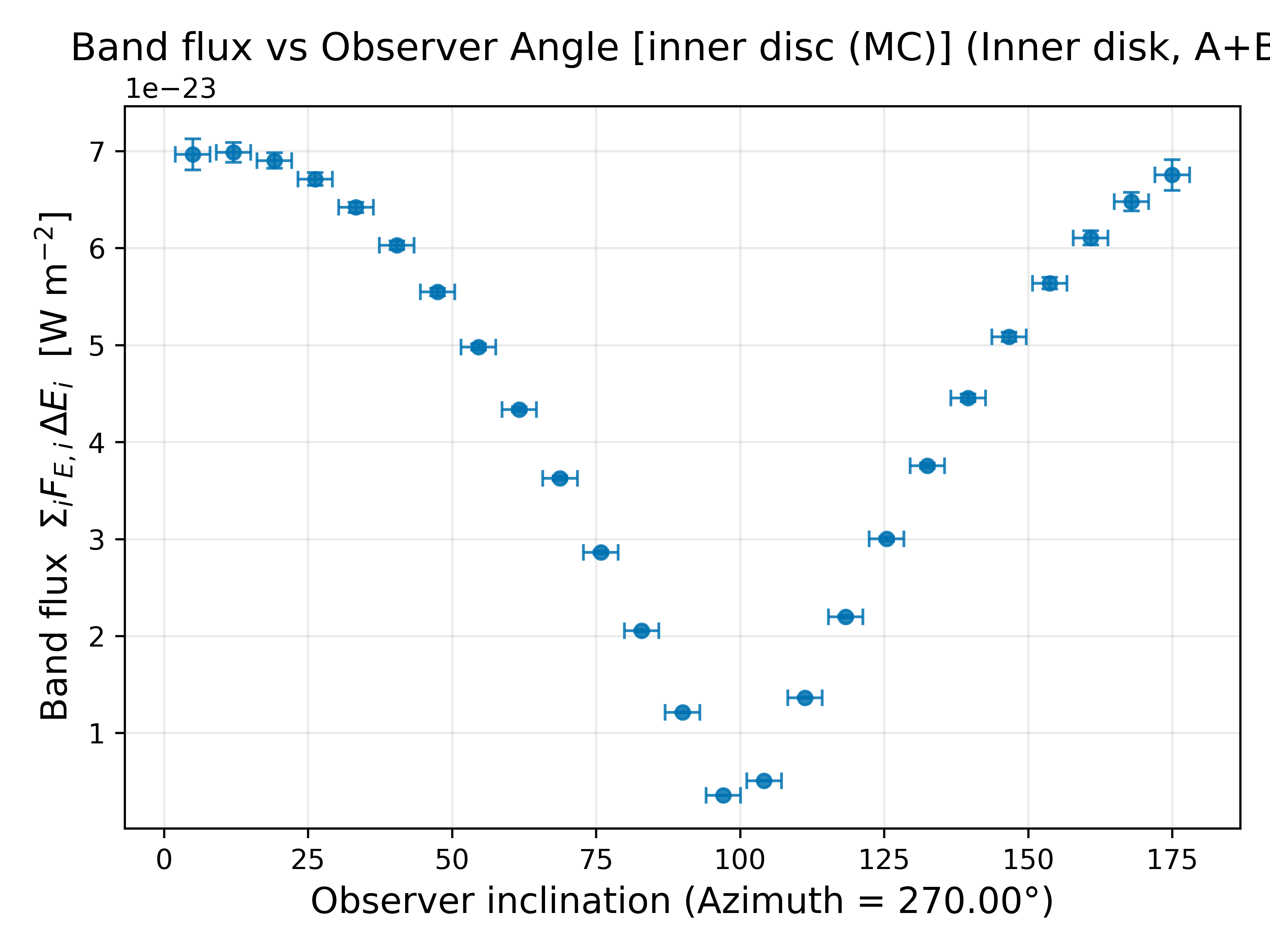}
    \caption{Flux density with inclination}
    \label{fig:Changing_inc_flux}
\end{subfigure}
\hfill
\begin{subfigure}{0.33\textwidth}
    \centering
    \includegraphics[width=\linewidth]{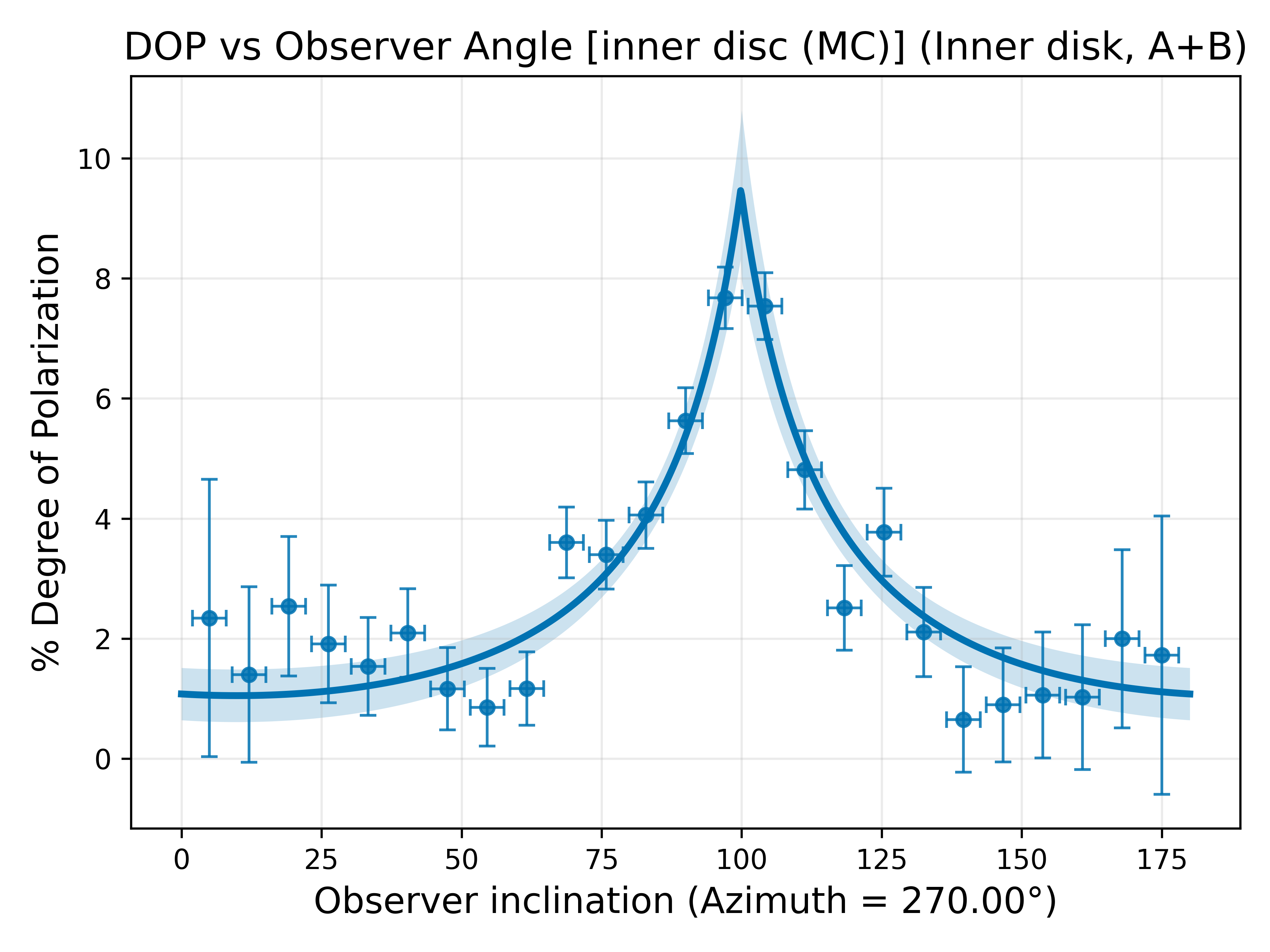}
    \caption{DOP with inclination}
    \label{fig:Changing_inc_DOP}
\end{subfigure}
\hfill
\begin{subfigure}{0.33\textwidth}
    \centering
    \includegraphics[width=\linewidth]{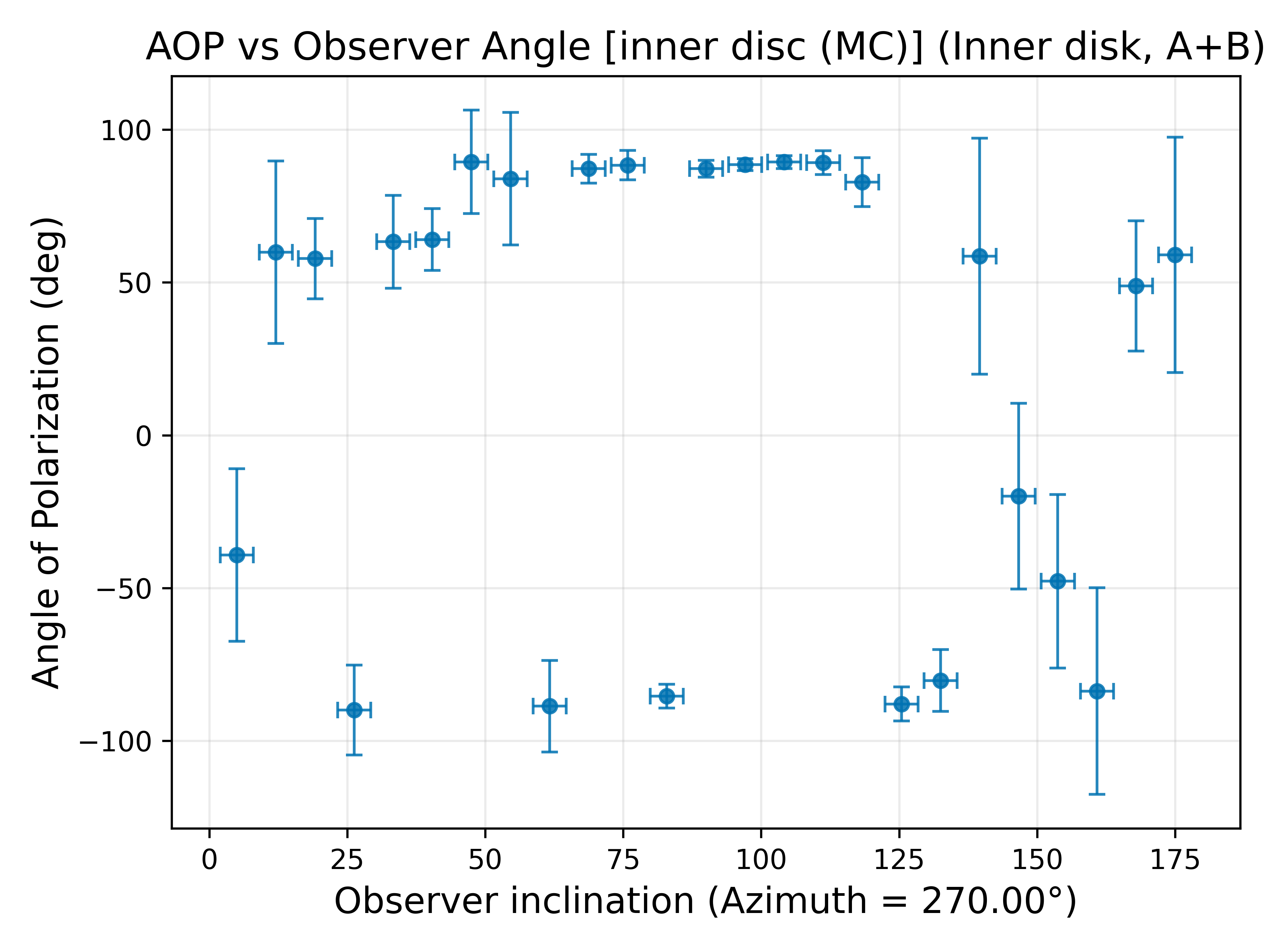}
    \caption{AOP with inclination}
    \label{fig:Changing_inc_AOP}
\end{subfigure}
\caption{Dependence of the flux density, DOP, and AOP of the inner-disc (Monte Carlo) component on observer inclination, for photons emitted from $R_0=9.5\,\Rg$ with outer-disc albedo $\rho_{\rm alb}=0.15$. The observer is at an azimuthal angle of $270^{\circ}$ with respect to the $X$ axis of the system, with a window of $\Delta\theta\times\Delta\phi=\pm3^\circ\times\pm3^\circ$. The curve in (b) is a fit of the form of Eq.~\eqref{eq:fit_incl} with $A=8.43\pm0.59\,\%$, $B=9.9\pm0.7^{\circ}$, $C=1.05\pm0.22\,\%$.}
\label{fig:changing_inclination_effects}
\end{figure*}

We next investigate how the polarization signature depends on the inclination of the observer. The polarization properties of the observed radiation depend not only on the intrinsic properties of the corona and disc, but also on the direction from which the system is viewed. Figure~\ref{fig:changing_inclination_effects} shows the variation of the weighted flux density (Figure~\ref{fig:Changing_inc_flux}), DOP (Figure~\ref{fig:Changing_inc_DOP}), and AOP (Figure~\ref{fig:Changing_inc_AOP}) as a function of observer inclination. The DOP peaks at approximately $100^{\circ}$, consistent with an edge-on view of the tilted inner disc. The offset $B$ in Eq.~\eqref{eq:fit_incl} therefore equals to the inner-disc tilt, and is fixed to that value for most of the cases.

\subsection{Dependence on coronal optical depth (electron density), without light bending}
\label{sec:corona_density_dependence}

Before turning to the light-bending-related parameter dependence below, we first isolate the effect of the corona's own scattering opacity, independent of any gravitational effect. The coronal electron density profile (Eq.~\ref{eq:elec_density}) sets the local Thomson mean free path $\bar\lambda(r) = 1/[n_e(r)\sigma_T]$, and therefore fixes how many Compton scatterings a photon typically undergoes before escaping the corona. Increasing the normalization $n_{{\rm e},0}$ increases the coronal optical depth at fixed corona radius $R_{\rm c}$, without changing the disc--corona geometry or the disc tilts. These runs are performed in flat spacetime, so that no gravitational effect is involved. This parameter therefore isolates the effect of \emph{scattering multiplicity} on the polarization signal, as distinct from the geometric and relativistic effects examined in the rest of this section.

\begin{figure*}
\centering
\begin{subfigure}{0.32\textwidth}
    \centering
    \includegraphics[width=\linewidth]{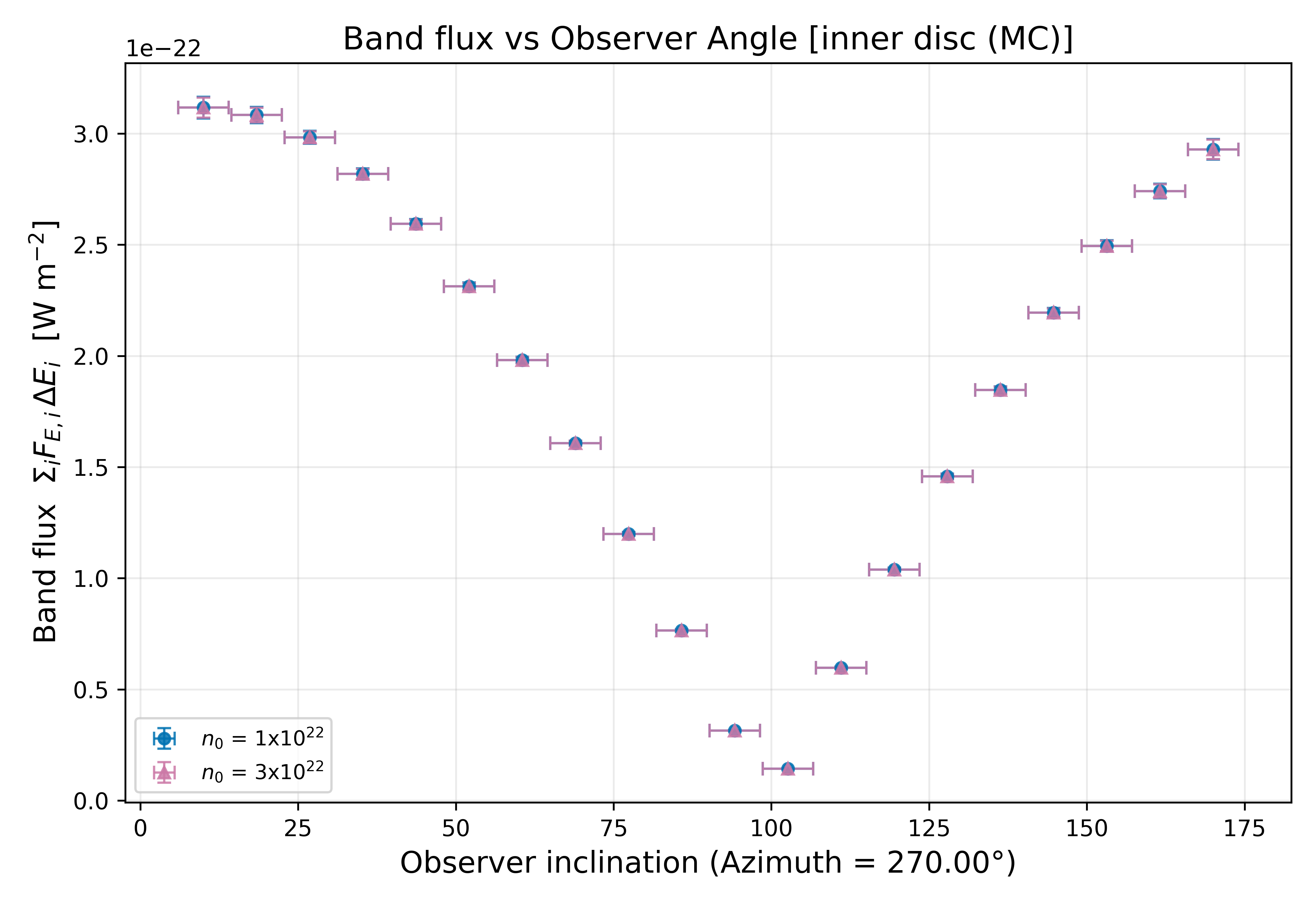}
    \caption{Flux density}
    \label{fig:flux_ne0}
\end{subfigure}
\hfill
\begin{subfigure}{0.32\textwidth}
    \centering
    \includegraphics[width=\linewidth]{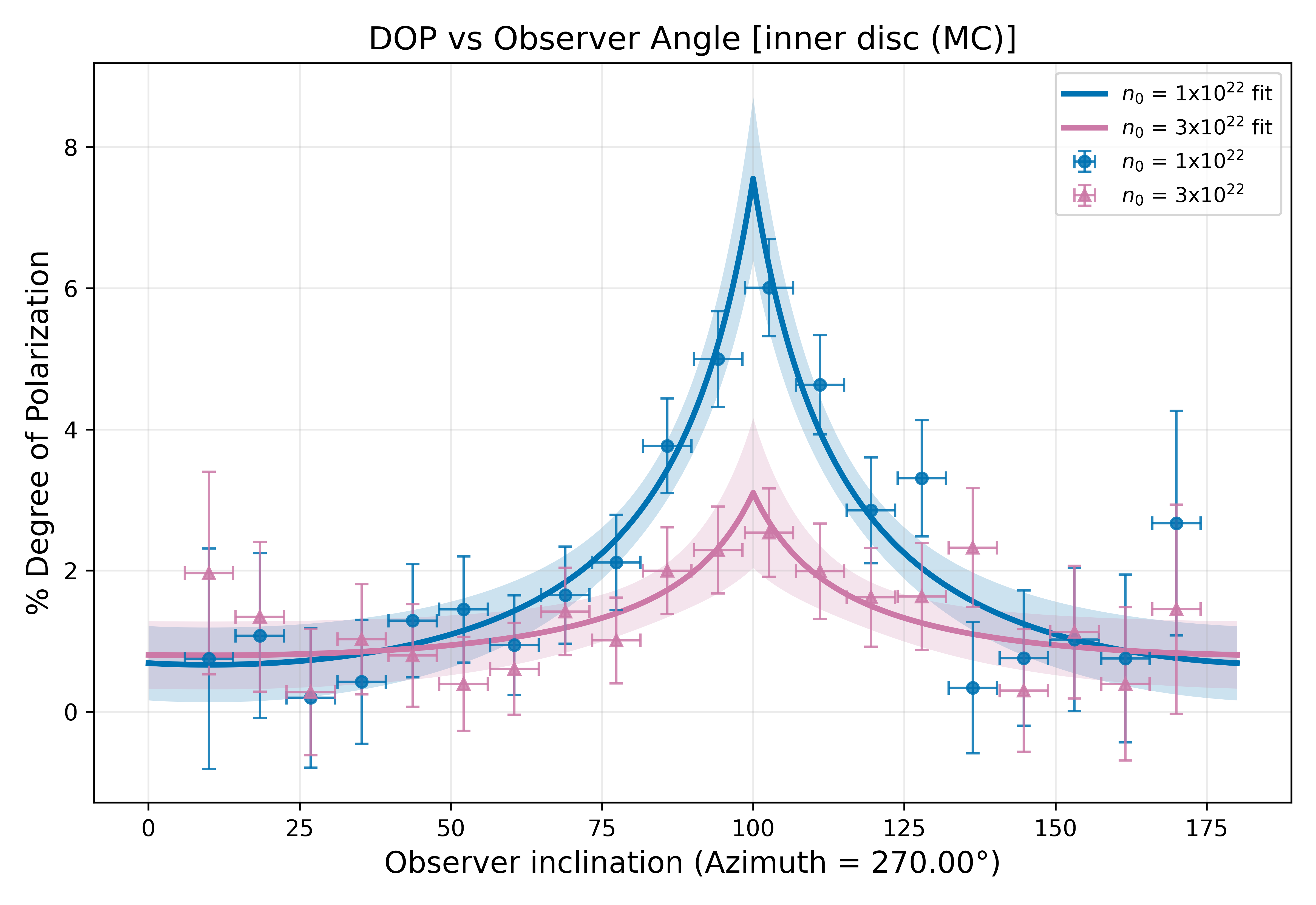}
    \caption{DOP}
    \label{fig:DOP_ne0}
\end{subfigure}
\hfill
\begin{subfigure}{0.32\textwidth}
    \centering
    \includegraphics[width=\linewidth]{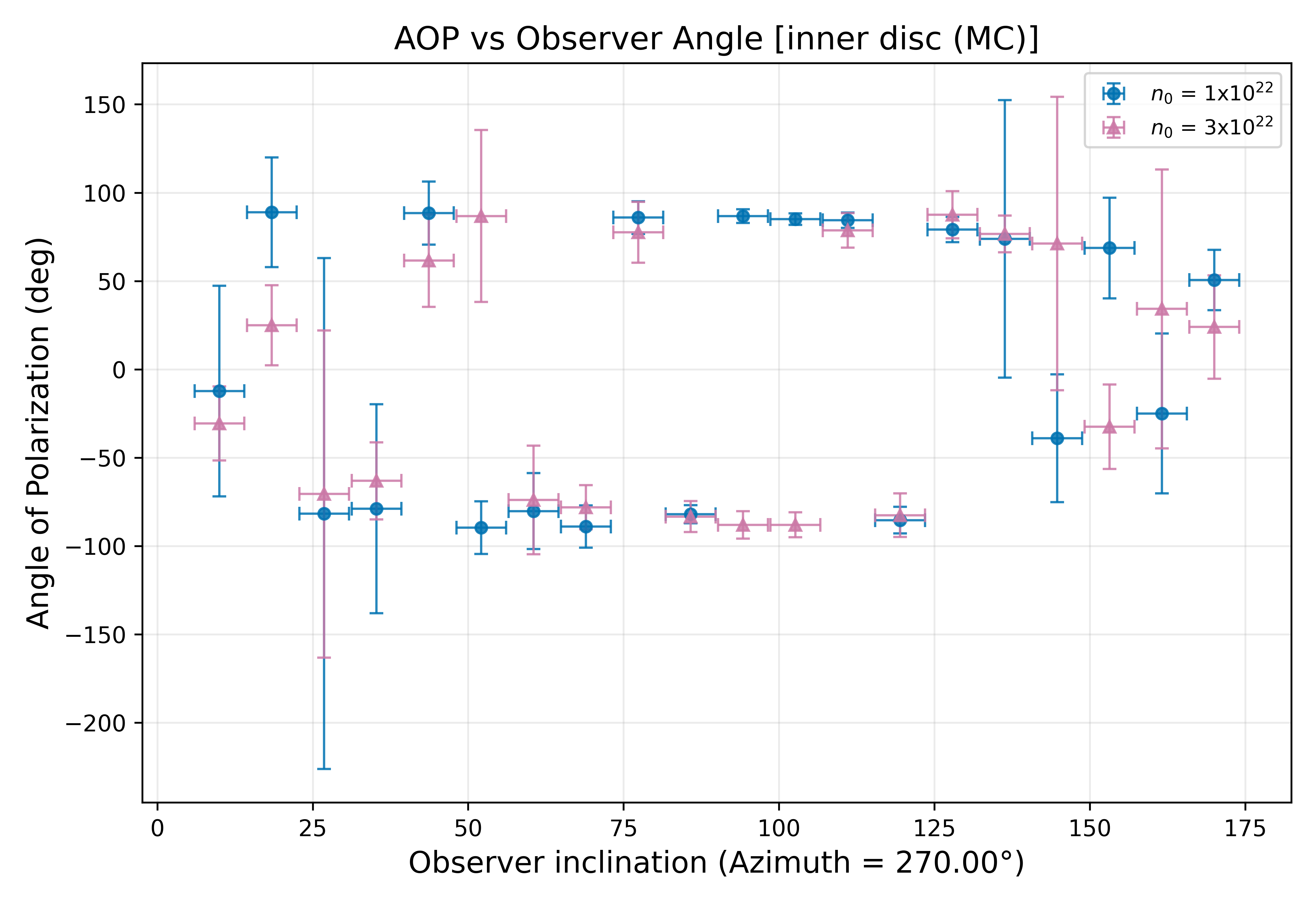}
    \caption{AOP}
    \label{fig:AOP_ne0}
\end{subfigure}
\caption{Dependence of the flux density, DOP, and AOP of the inner-disc (Monte Carlo) component on observer inclination for two values of the coronal electron-density normalization, $n_{{\rm e},0}=1\times10^{22}$ and $3\times10^{22}\,\mathrm{m^{-3}}$, at fixed corona radius $R_{\rm c}$ and disc geometry. The outer-disc albedo is $\rho_{\rm alb}=0.2$ and the observer is at an azimuthal angle of $270^{\circ}$ with a window of $\Delta\theta\times\Delta\phi=\pm4^\circ\times\pm4^\circ$. The curves in (b) are fits of the form of Eq.~\eqref{eq:fit_incl}, with $A=6.89\pm0.76\,\%$, $B=10.0^{\circ}$, $C=0.66\pm0.27\,\%$ for $n_{{\rm e},0}=1\times10^{22}\,\mathrm{m^{-3}}$ and with $A=2.30\pm0.69\,\%$, $B=10.0^{\circ}$, $C=0.80\pm0.25\,\%$ for $n_{{\rm e},0}=3\times10^{22}\,\mathrm{m^{-3}}$. Increasing $n_{{\rm e},0}$ increases the typical number of Compton scatterings per escaping photon and systematically reduces the DOP.}
\label{fig:corona_density_dependence}
\end{figure*}

We compare two representative values of the density normalization, $n_{{\rm e},0} = 1\times10^{22}\,\mathrm{m^{-3}}$ and $n_{{\rm e},0} = 3\times10^{22}\,\mathrm{m^{-3}}$, holding all other system parameters fixed at the fiducial values of Section~\ref{sec:fiducial} and with the photons propagating along straight lines in flat spacetime. Figure~\ref{fig:corona_density_dependence} shows the resulting flux, DOP, and AOP as functions of observer inclination for both cases.

We find that the higher-density corona produces a systematically lower DOP across the range of inclinations (Figure~\ref{fig:DOP_ne0}). This is consistent with the higher optical depth driving a larger typical number of Compton scatterings per escaping photon. Each Thomson scattering event partially reorients and depolarizes the photon's polarization vector relative to its state before that scattering, so as the number of scatterings accumulated along a photon's path increases, the net polarization retained by the escaping ensemble is increasingly averaged down. Quantitatively, the fitted amplitude $A$ of Eq.~\eqref{eq:fit_incl} falls from $11.0\pm0.6\%$ with scattering switched off (Section~\ref{sec:light_bending_only}) to $6.9\pm0.8\%$ for $n_{{\rm e},0}=1\times10^{22}\,\mathrm{m^{-3}}$ and $2.3\pm0.7\%$ for $3\times10^{22}\,\mathrm{m^{-3}}$ (These runs differ only in the observer window and value of azimuth -- see the figure captions).This is the same qualitative depolarizing mechanism that underlies the saturation of the emergent polarization degree with increasing optical depth in the classical semi-infinite-atmosphere (Milne-problem) limit \citep{Chandrasekhar1960}. Here it is seen in a finite, spherical corona with varying optical depth, rather than in a plane-parallel slab. The AOP (Figure~\ref{fig:AOP_ne0}) correspondingly shows a larger scatter about its expected value ($\pm 90^\circ$ for this observer position) in the higher-density case, mainly at the extremes of the inclination range. A larger number of individually reorienting scattering events increases the variance of the final polarization angle even where the ensemble mean is preserved by the symmetry of the system. 

\subsection{Dependence on light bending alone, without scattering}
\label{sec:light_bending_only}

Having isolated the effect of scattering multiplicity in Section~\ref{sec:corona_density_dependence}, we now isolate the opposite limit: the effect of gravitational light bending on the observed polarization, with all Compton scattering switched off. In this configuration the photons emitted from the disc propagate directly to the observer, so any difference between the flat-spacetime and BH cases arises purely from the curved geometry. This allows us to separate the purely relativistic contribution to the DOP and AOP from the scattering-induced contribution discussed above.

\begin{figure*}
\centering
\begin{subfigure}{0.32\textwidth}
    \centering
    \includegraphics[width=\linewidth]{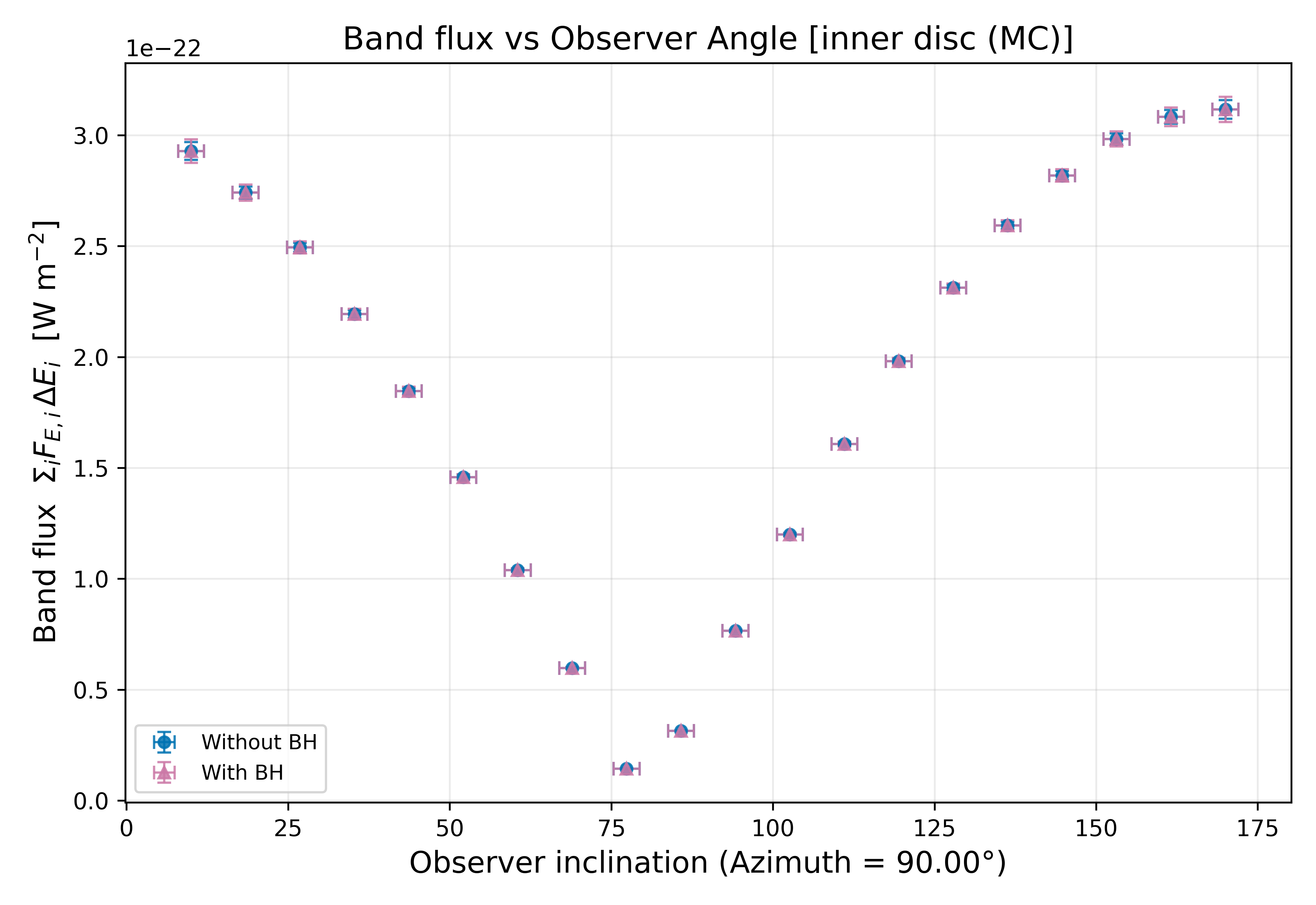}
    \caption{Flux density}
    \label{fig:flux_lb}
\end{subfigure}
\hfill
\begin{subfigure}{0.32\textwidth}
    \centering
    \includegraphics[width=\linewidth]{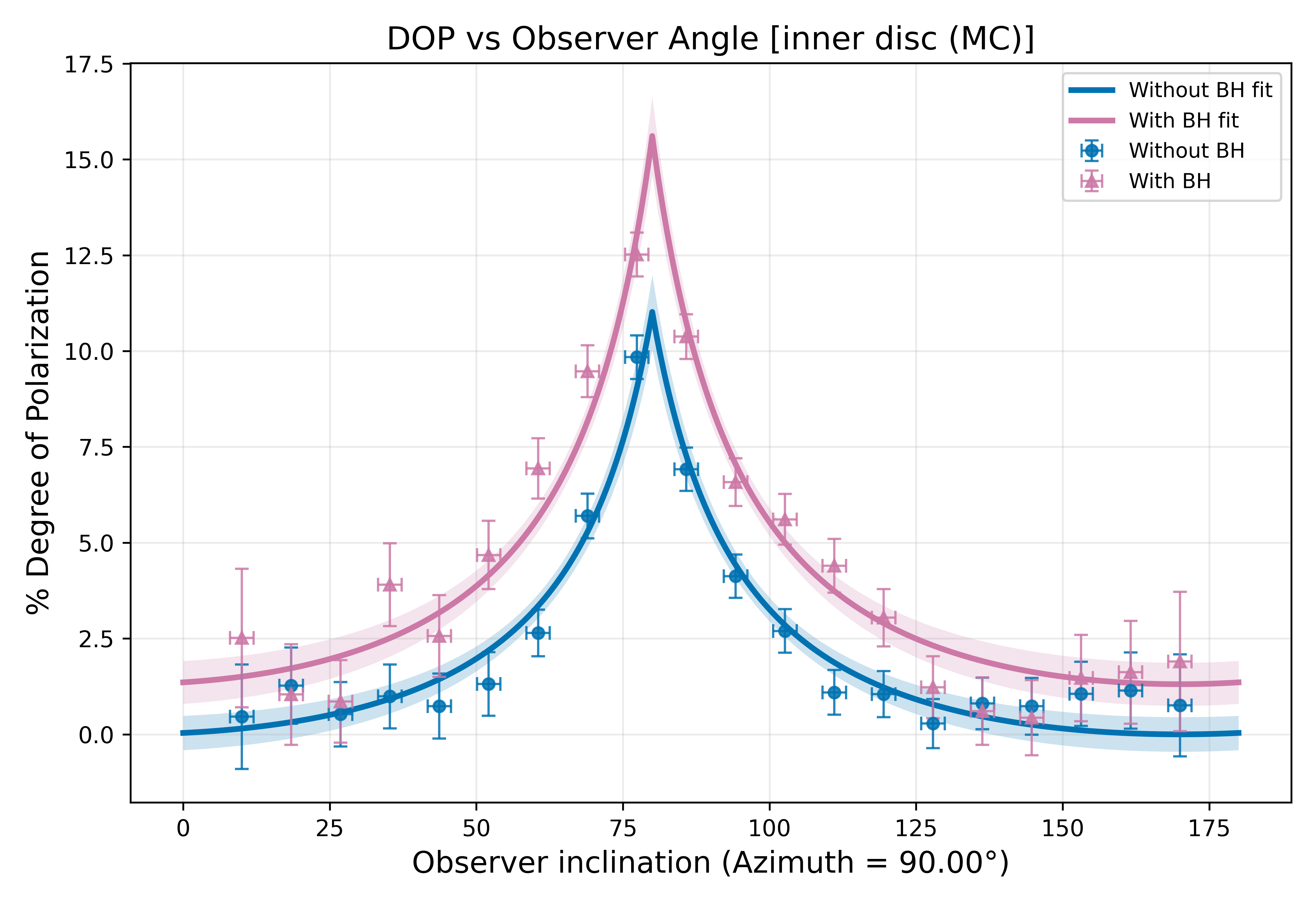}
    \caption{DOP}
    \label{fig:DOP_lb}
\end{subfigure}
\hfill
\begin{subfigure}{0.32\textwidth}
    \centering
    \includegraphics[width=\linewidth]{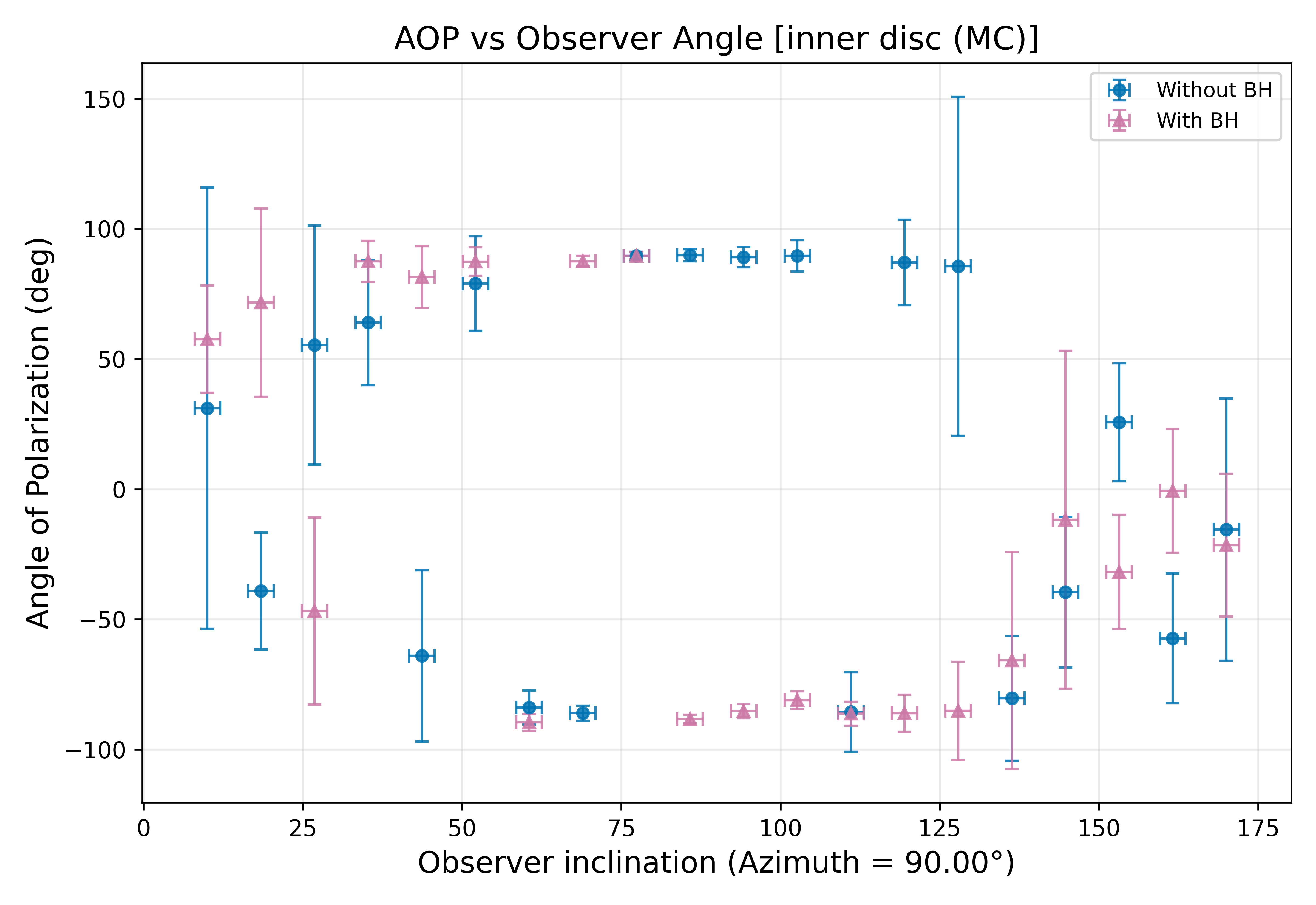}
    \caption{AOP}
    \label{fig:AOP_lb}
\end{subfigure}
\caption{Dependence of the flux density, DOP, and AOP of the inner-disc (Monte Carlo) component on observer inclination, with and without gravitational light bending (flat spacetime versus Kerr BH) and with no Compton scattering. The outer-disc albedo is $\rho_{\rm alb}=0.2$ and the observer is at an azimuthal angle of $90^{\circ}$ with a window of $\Delta\theta\times\Delta\phi=\pm2^\circ\times\pm2^\circ$. The curves in (b) are fits of the form of Eq.~\eqref{eq:fit_incl}, with $A=11.02\pm0.64\,\%$, $B=-10.0^{\circ}$, $C=0.00\pm0.23\,\%$ without light bending and with $A=14.30\pm0.71\,\%$, $B=-10.0^{\circ}$, $C=1.31\pm0.29\,\%$ with light bending. The flux is unchanged by the presence of the BH, whereas the DOP increases because photons near the equatorial plane of the BH are captured, removing strongly bent, lower-DOP trajectories from the observed ensemble.}
\label{fig:light_bending_only}
\end{figure*}

We compare two cases: (i) photons propagating along straight lines in flat spacetime, and (ii) photons propagating along null geodesics in the Kerr spacetime of the BH, with all other system parameters held fixed at the fiducial values of Section~\ref{sec:fiducial}. Figure~\ref{fig:light_bending_only} shows the resulting flux, DOP, and AOP as functions of observer inclination for both cases.

The observed flux is identical in the two cases (Figure~\ref{fig:flux_lb}), since without scattering the total number of photons reaching the observer from the emitting region is unchanged at the level of our simulation. The DOP, however, is systematically higher in the presence of the BH (Figure~\ref{fig:DOP_lb}). This can be understood as follows. In the flat-spacetime case, the escaping ensemble includes photons that leave the disc towards the equatorial plane of the BH. In the BH case, many of these photons are engulfed by the BH, while those that survive are strongly deflected, and their trajectories are bent so that they reach the observer with a polarization state that has been rotated relative to the initial one. These light-bent trajectories carry a lower DOP, and when averaged with the less-deflected photons they reduce the net polarization of the ensemble. Removing a significant fraction of them through capture by the BH therefore leaves an ensemble dominated by less-deflected photons, and the overall DOP rises.

This result has an important consequence for the interpretation of the scattering-free baseline. Even if we adopt as the starting point the maximum intrinsic DOP of $11.71\%$ from the classical Chandrasekhar result for a scattering atmosphere \citep{Chandrasekhar1960}, merely introducing a BH raises the DOP seen by the observer, as shown in Figure~\ref{fig:DOP_lb}. The enhancement is therefore not a consequence of scattering. It is a purely geometric effect of the BH removing the most strongly bent, and hence most depolarized, photon paths from the observed ensemble. This contrasts with the behaviour in Section~\ref{sec:corona_density_dependence}, where the dominant effect was the depolarization caused by multiple scatterings.

The AOP (Figure~\ref{fig:AOP_lb}) behaves as in the previous sections. For most inclinations it remains at the expected value of $\pm 90^\circ$ for this observer position, while at the extremes of the inclination range it deviates from this value. As before, the ensemble mean is preserved by the symmetry of the system, but the number of photons reaching the observer from these extreme viewing angles is small, and the strongly bent trajectories there reorient the polarization vector by different amounts, which increases the scatter of the final polarization angle.

\subsection{Parameter dependence of the inclination signature: reflection fraction and light bending}
\label{sec:parameter_dependence_inclination}

Having established the dependence of the polarization signatures on the observer inclination, and the independent contributions of scattering and of the BH (light bending), we next investigate how the inclination-dependent DOP, AOP, and flux are modified by changes in the physical parameters of the system. The observer inclination is varied while the effects of the reflection fraction and of gravitational effects are examined separately, by varying the albedo and the photon-emission radius, respectively.

\begin{figure*}
\centering
\begin{subfigure}{0.32\textwidth}
    \centering
    \includegraphics[width=\linewidth]{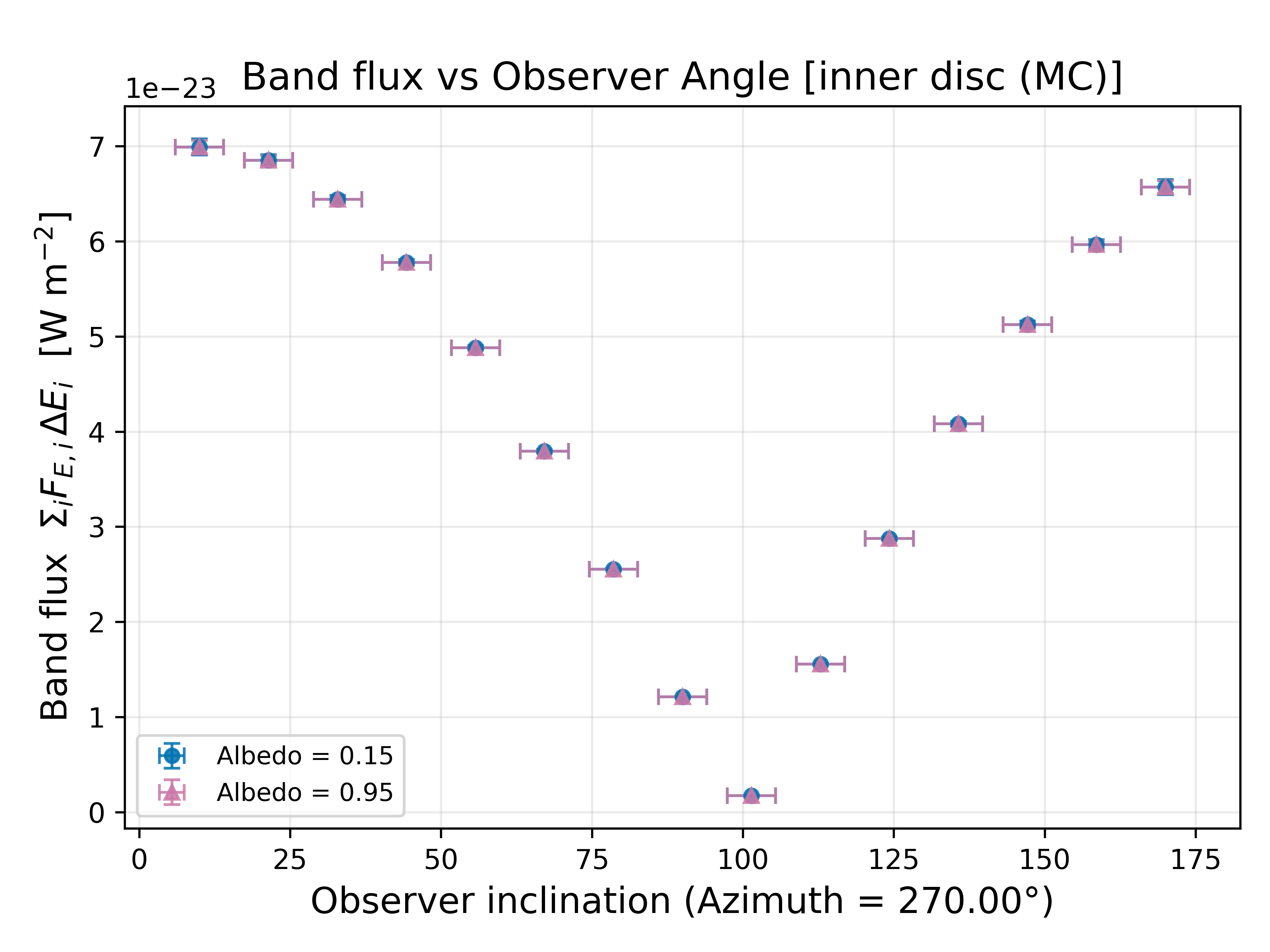}
    \caption{Flux density}
    \label{fig:Inc_albedo_flux}
\end{subfigure}
\hfill
\begin{subfigure}{0.32\textwidth}
    \centering
    \includegraphics[width=\linewidth]{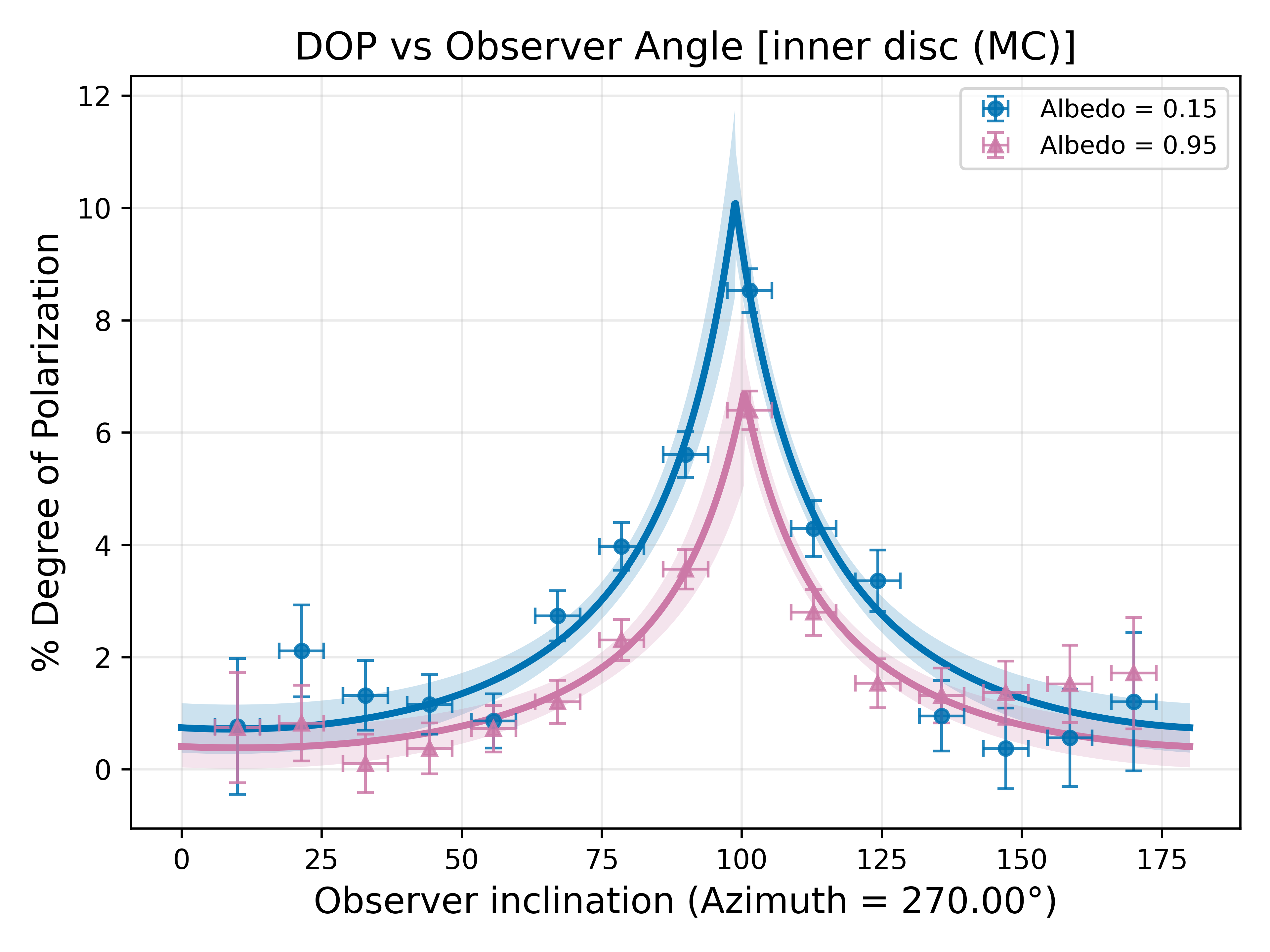}
    \caption{DOP}
    \label{fig:Inc_albedo_DOP}
\end{subfigure}
\hfill
\begin{subfigure}{0.32\textwidth}
    \centering
    \includegraphics[width=\linewidth]{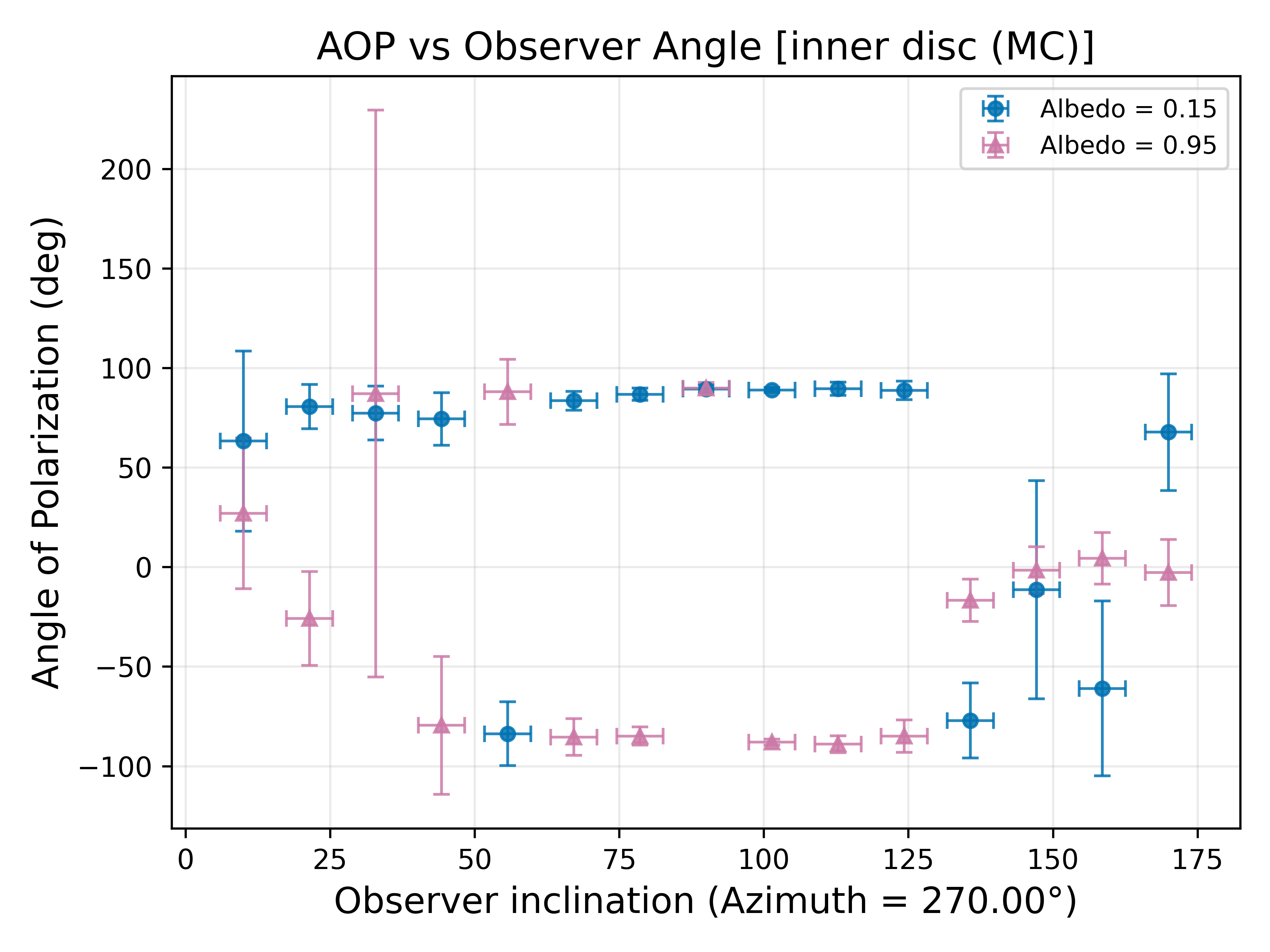}
    \caption{AOP}
    \label{fig:Inc_albedo_AOP}
\end{subfigure}
\caption{Dependence of the flux density, DOP, and AOP of the inner-disc (Monte Carlo) component on observer inclination for different values of the outer-disc albedo, $\rho_{\rm alb}=0.15$ and $0.95$. Photons are emitted from $R_0=9.5\,\Rg$, and the observer is at an azimuthal angle of $270^{\circ}$ with a window of $\Delta\theta\times\Delta\phi=\pm4^\circ\times\pm4^\circ$. The curves in (b) are fits of the form of Eq.~\eqref{eq:fit_incl}, with $A=9.42\pm0.63\,\%$, $B=8.8\pm0.7^{\circ}$, $C=0.72\pm0.22\,\%$ for $\rho_{\rm alb}=0.15$ and with $A=6.34\pm0.57\,\%$, $B=10.5\pm0.9^{\circ}$, $C=0.38\pm0.19\,\%$ for $\rho_{\rm alb}=0.95$. The flux is unaffected by the albedo. A higher albedo lowers the DOP, makes the DOP profile asymmetric about its peak, and increases the scatter of the AOP about $90^{\circ}$ at extreme inclinations.}
\label{fig:inclination_albedo_dependence}
\end{figure*}

The effect of the reflection fraction is investigated by varying the albedo of the outer disc. Since the albedo determines the probability that photons incident on the outer disc are reflected rather than absorbed, changing this parameter modifies the relative contribution of reflected photons to the observed radiation. The corresponding results are shown in Figure~\ref{fig:inclination_albedo_dependence}.

As shown in Figure~\ref{fig:Inc_albedo_flux}, changing the albedo has no discernible effect on the observed modulation of the flux density, since the reflected photons carry substantially lower weights than the remaining photon population, as shown in Paper~I. However, Figure~\ref{fig:Inc_albedo_DOP} shows a clear decrease in the DOP for the system with the higher albedo. It also shows an asymmetry in the way the DOP approaches and departs from its peak value on either side of the peak, and this difference is most prominent in the range of inclinations expected to be affected most by reflected photons. This confirms that the reflection fraction of photons from the outer disc does influence the observed DOP. The effect arises because the reflection from the outer disc is modelled as Compton scattering, which is known to depolarize the radiation. As a larger fraction of the photons is depolarized by reflection and redistributed throughout the system, the overall DOP drops. Figure~\ref{fig:Inc_albedo_AOP} shows the corresponding variation of the AOP. For the lower albedo it mostly remains close to $\sim\!90^{\circ}$ for the corresponding observer, whereas for the higher albedo it shows a considerable modulation, mainly at extreme inclinations.

\begin{figure*}
\centering
\begin{subfigure}{0.32\textwidth}
    \centering
    \includegraphics[width=\linewidth]{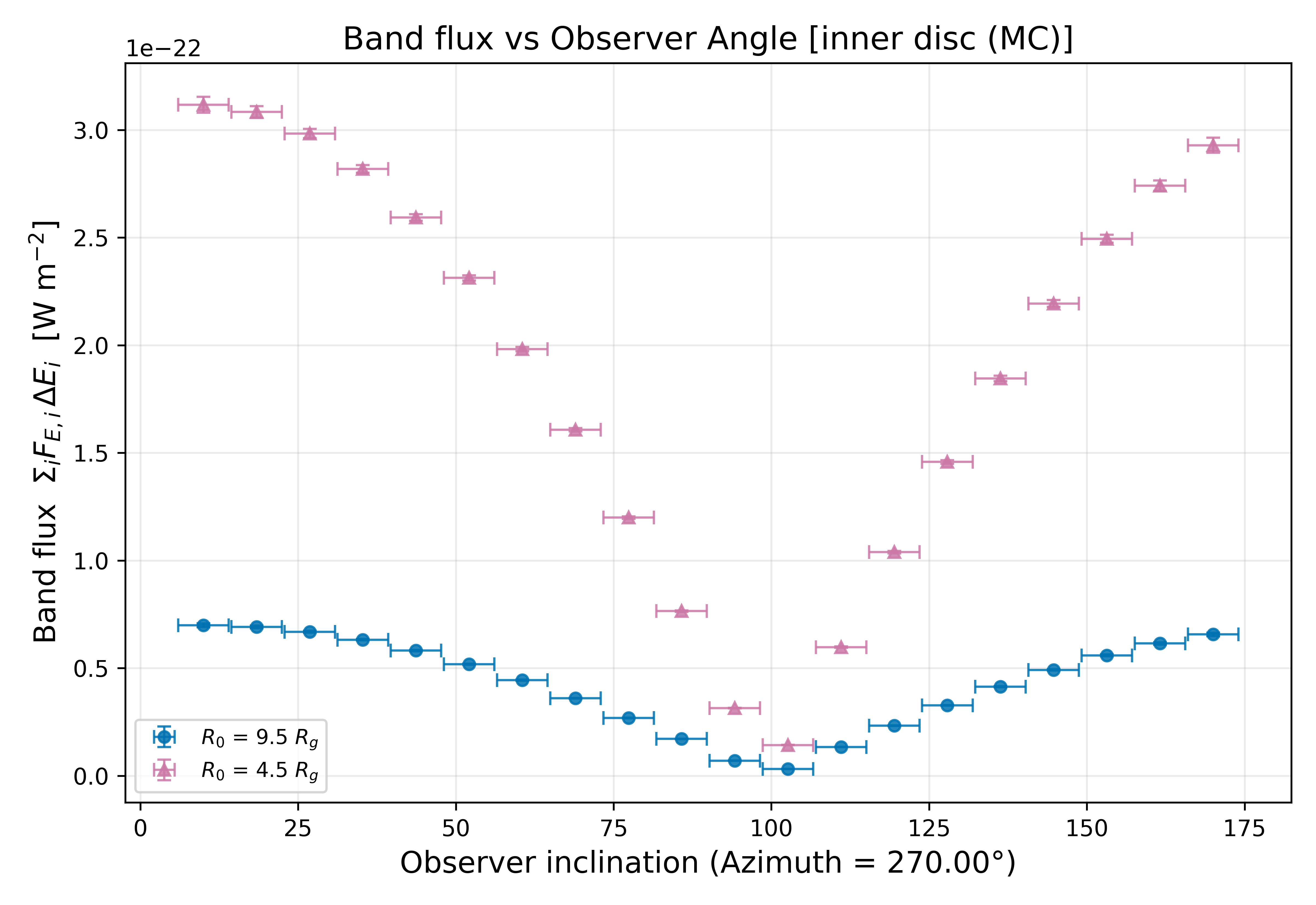}
    \caption{Flux density}
    \label{fig:Inc_emission_radius_flux}
\end{subfigure}
\hfill
\begin{subfigure}{0.32\textwidth}
    \centering
    \includegraphics[width=\linewidth]{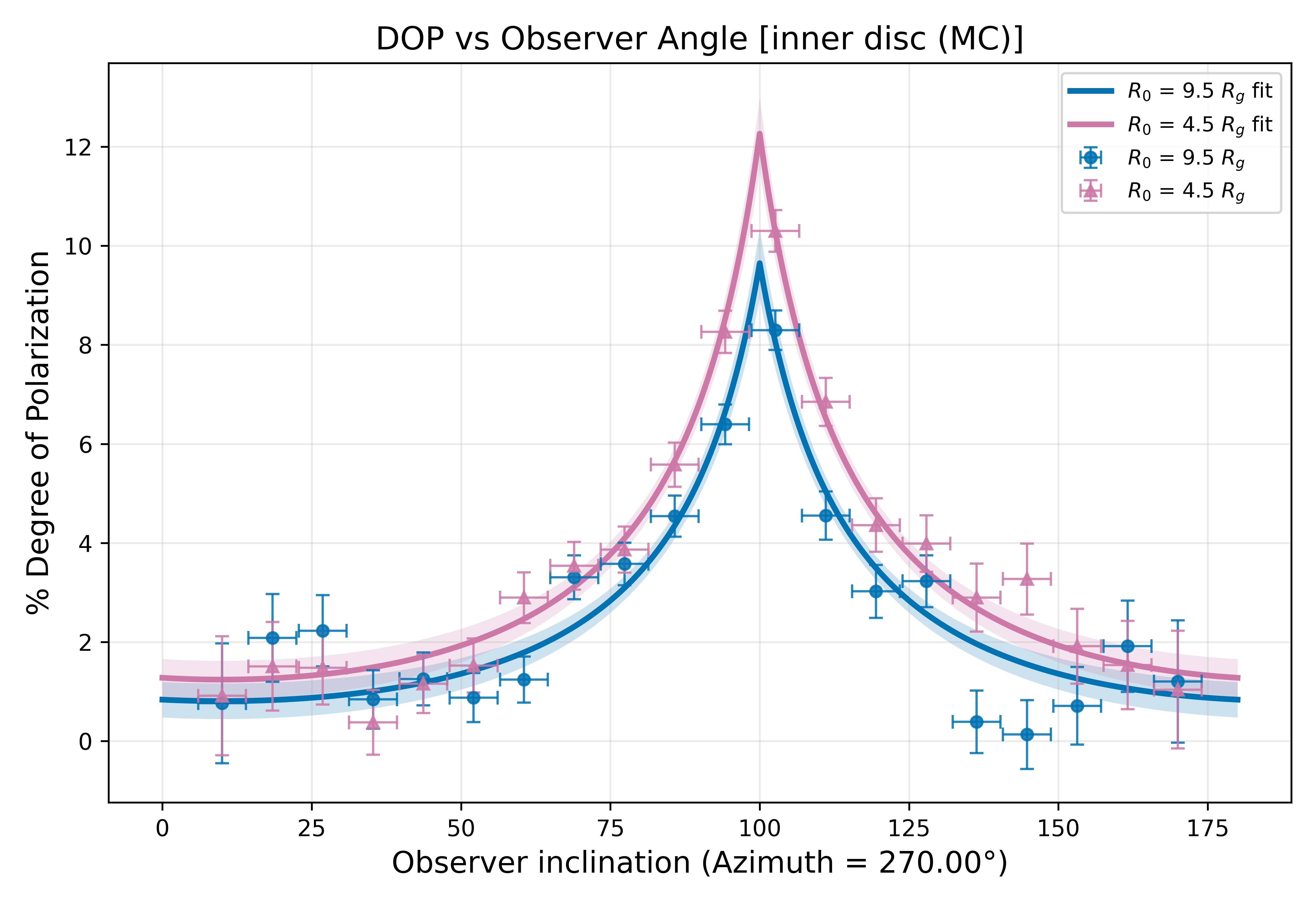}
    \caption{DOP}
    \label{fig:Inc_emission_radius_DOP}
\end{subfigure}
\hfill
\begin{subfigure}{0.32\textwidth}
    \centering
    \includegraphics[width=\linewidth]{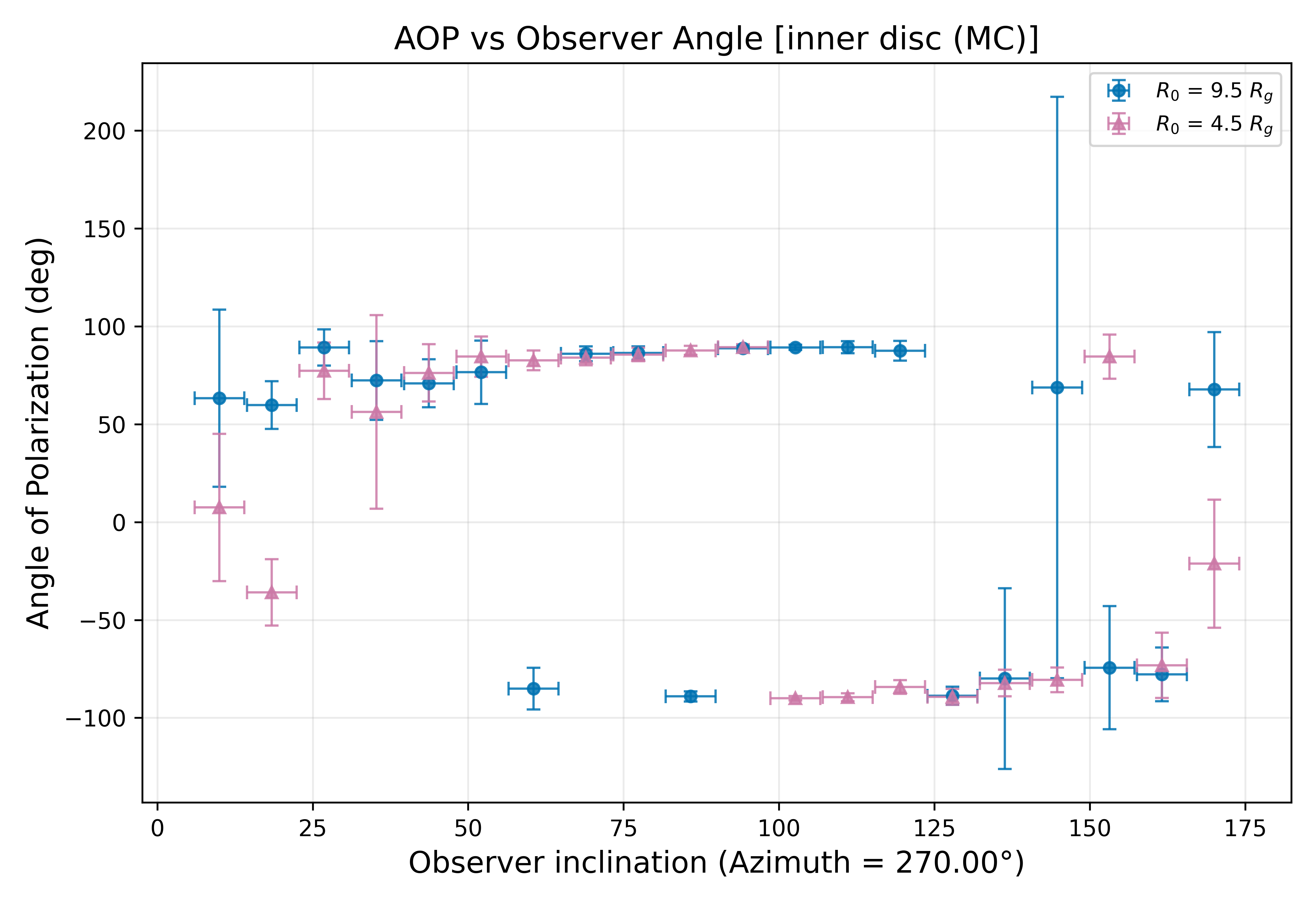}
    \caption{AOP}
    \label{fig:Inc_emission_radius_AOP}
\end{subfigure}
\caption{Dependence of the flux density, DOP, and AOP of the inner-disc (Monte Carlo) component on observer inclination for different photon-emission radii, $R_0=9.5\,\Rg$ and $4.5\,\Rg$. The outer-disc albedo is $\rho_{\rm alb}=0.15$ and the observer is at an azimuthal angle of $270^{\circ}$ with a window of $\Delta\theta\times\Delta\phi=\pm4^\circ\times\pm4^\circ$. The curves in (b) are fits of the form of Eq.~\eqref{eq:fit_incl}, with $A=8.84\pm0.48\,\%$, $B=10.0^{\circ}$, $C=0.81\pm0.18\,\%$ for $R_0=9.5\,\Rg$ and with $A=11.02\pm0.50\,\%$, $B=10.0^{\circ}$, $C=1.24\pm0.19\,\%$ for $R_0=4.5\,\Rg$. The flux differs between the two radii because the seed-photon temperature, and hence the normalization of the inner-disc component relative to the outer disc, depends on $R_0$. A smaller $R_0$ corresponds to stronger gravitational light bending, which raises the DOP and increases the scatter of the AOP about $90^{\circ}$ at extreme inclinations.}
\label{fig:inclination_emission_radius_dependence}
\end{figure*}

We further examine the influence of gravitational effects, in the presence of coronal Compton scattering, by varying the photon-emission radius while studying the polarization as a function of observer inclination. Changing the emission radius modifies the region from which the photons originate, and consequently the extent to which the photon trajectories are affected by the strong gravitational field. The resulting changes in the flux, DOP, and AOP with observer inclination are shown in Figure~\ref{fig:inclination_emission_radius_dependence}. The flux differs between the two emission radii, as expected (Figure~\ref{fig:Inc_emission_radius_flux}): the temperature of the seed photons launched from the two radii is different, and so the normalization of the inner-disc flux (\texttt{mc\_scale}) relative to the outer disc is also different. For the DOP (Figure~\ref{fig:Inc_emission_radius_DOP}) we observe the same trend as in Section~\ref{sec:light_bending_only}, where we explained why the DOP increases with increasing light-bending effects. In the present case this effect is superposed on the depolarization caused by coronal scattering, but the overall behaviour still holds: when photons are emitted from a radius closer to the ISCO, and are therefore more strongly affected by light bending, the DOP is higher than when they are emitted further from the ISCO. The AOP (Figure~\ref{fig:Inc_emission_radius_AOP}) behaves as in the previous cases. For weaker light bending it mostly remains close to $\sim\!90^{\circ}$ for the specific observer, whereas for stronger light bending it shows a considerable modulation at extreme inclinations.
These comparisons allow the inclination-dependent polarization signatures to be examined in the presence of different reflection fractions and different strengths of gravitational effects. The corresponding changes in the observed flux provide an additional measure of how these physical processes modify the radiation reaching the observer.
\subsection{Polarization from a precessing inner accretion disc}
\label{sec:precessing_disc}

We next consider the precession of the inner accretion disc and investigate the resulting polarization signatures. The changing orientation of the inner disc relative to the observer introduces a phase-dependent modulation of the observed polarization properties. We examine the resulting variations in the DOP and AOP as functions of precession phase.

\begin{figure*}
\centering
\begin{subfigure}{0.32\textwidth}
    \centering
    \includegraphics[width=\linewidth]{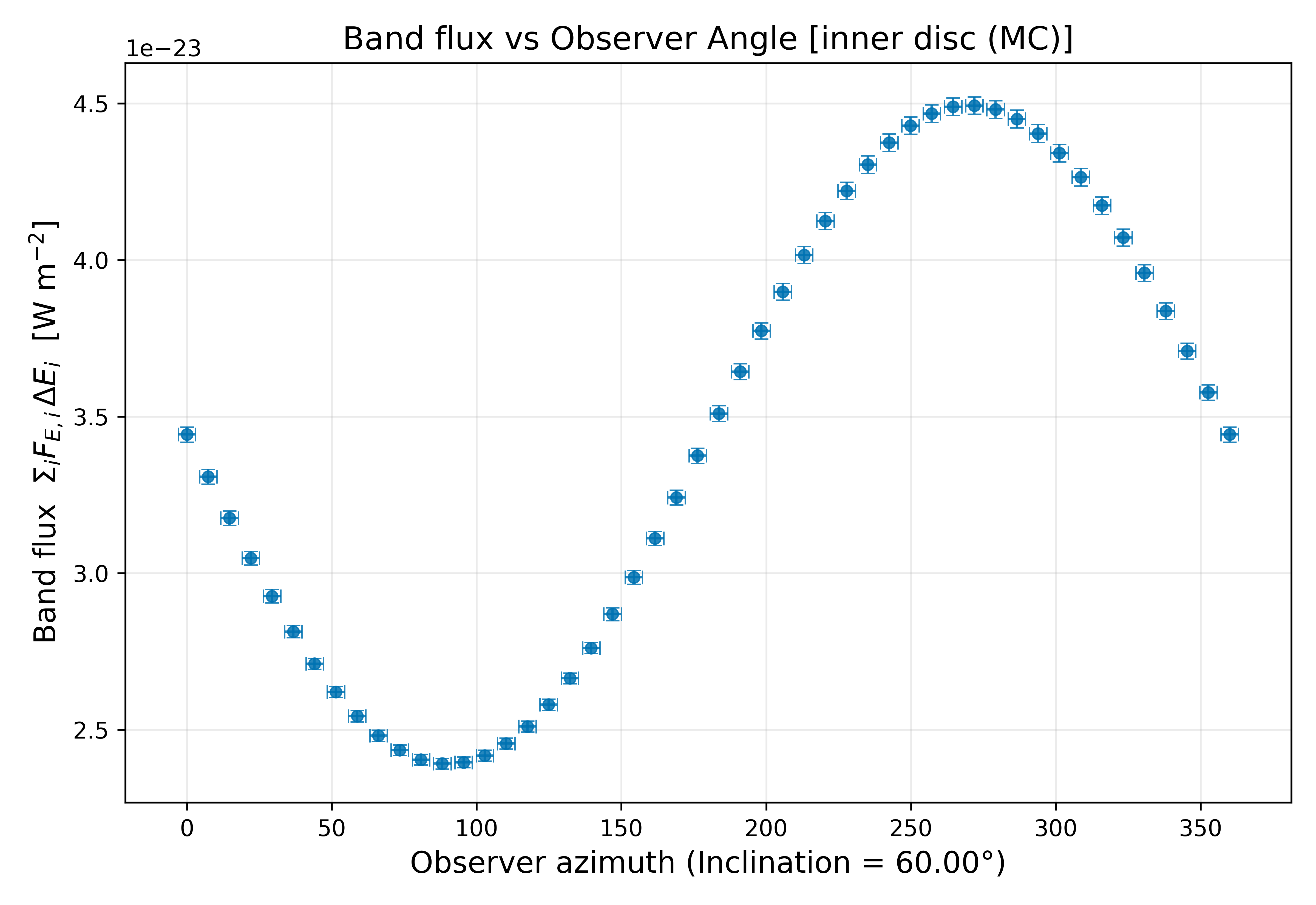}
    \caption{Flux density variation}
    \label{fig:precessing_flux}
\end{subfigure}
\hfill
\begin{subfigure}{0.32\textwidth}
    \centering
    \includegraphics[width=\linewidth]{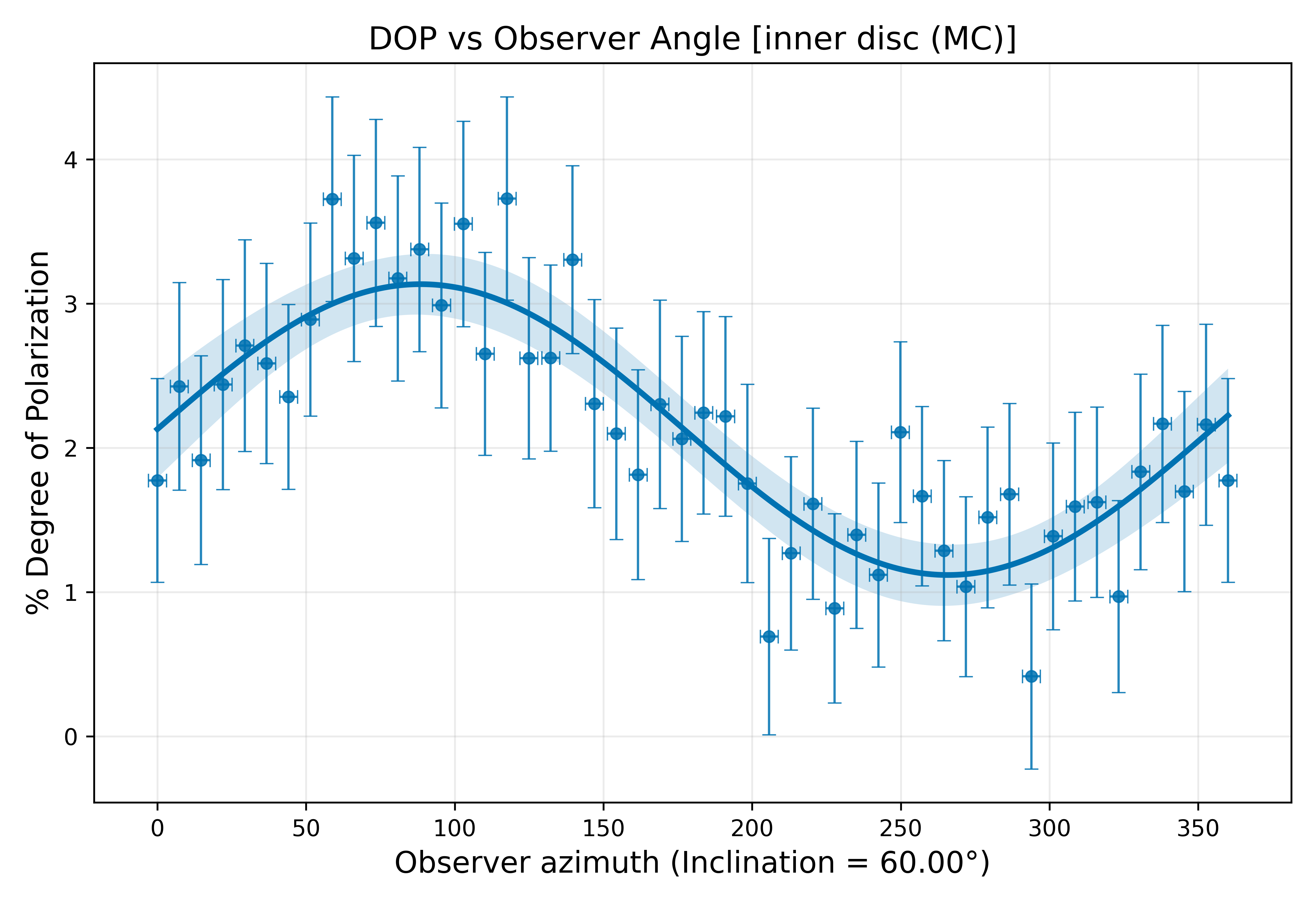}
    \caption{DOP variation}
    \label{fig:precessing_DOP}
\end{subfigure}
\hfill
\begin{subfigure}{0.32\textwidth}
    \centering
    \includegraphics[width=\linewidth]{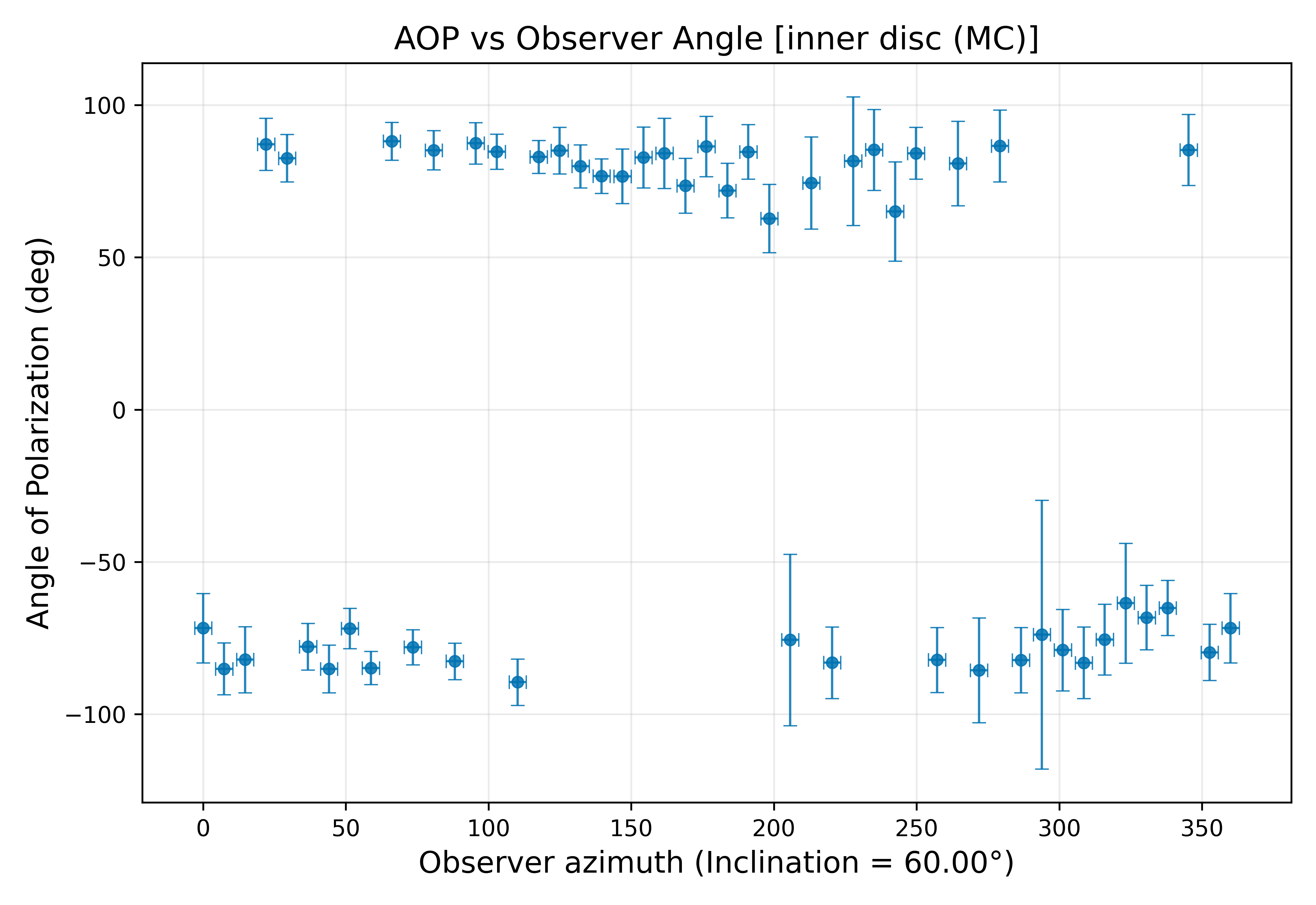}
    \caption{AOP variation}
    \label{fig:precessing_DOP_AOP}
\end{subfigure}
\caption{Variation of the flux density, DOP, and AOP of the inner-disc (Monte Carlo) component with precession phase, obtained by sweeping the observer azimuth $\phi_{\rm obs}$ to replicate the precession of the inner accretion disc. The observer is at an inclination of $60^{\circ}$ with respect to the BH spin axis, with a window of $\Delta\theta\times\Delta\phi=\pm3^\circ\times\pm3^\circ$, and the outer-disc albedo is $\rho_{\rm alb}=0.15$. Photons are emitted from $R_0=9.5\,\Rg$. The curve in (b) is a fit of the form of Eq.~\eqref{eq:fit_prec}, with $A=1.007\pm0.087\%$, $\omega=0.0177\pm0.0007$, $\varphi=0.003\pm0.156$, $C=2.126\pm0.059\%$. The DOP is modulated similarly to, but out of phase with, the flux, and the AOP deviates from $90^{\circ}$ over the cycle.}
\label{fig:precessing_effects}
\end{figure*}

The polarization signatures associated with the precession of the inner accretion disc are shown in Figure~\ref{fig:precessing_effects}. From Figures~\ref{fig:precessing_flux} and \ref{fig:precessing_DOP}, it is evident that the DOP is modulated in a manner similar to, but out of phase with, the modulation of the flux density. The conditions under which the DOP or AOP modulation may decrease or vanish are explored in the following subsection. Figure~\ref{fig:precessing_DOP_AOP} further shows that the AOP deviates from $\pm90^{\circ}$ over the precession cycle, and tends to bend inwards, towards the projection of the BH spin axis onto the image plane of the observer.

\subsection{Parameter dependence of the precessing polarization}
\label{sec:precession_parameter_dependence}

We now investigate how variations in the physical parameters of the model affect the polarization signatures in the presence of inner-disc precession. The changing orientation of the inner accretion disc provides a phase-dependent viewing geometry, and the DOP, AOP, and flux are examined as functions of the precession phase for different physical and geometric configurations.

\begin{figure*}
\centering
\begin{subfigure}{0.32\textwidth}
    \centering
    \includegraphics[width=\linewidth]{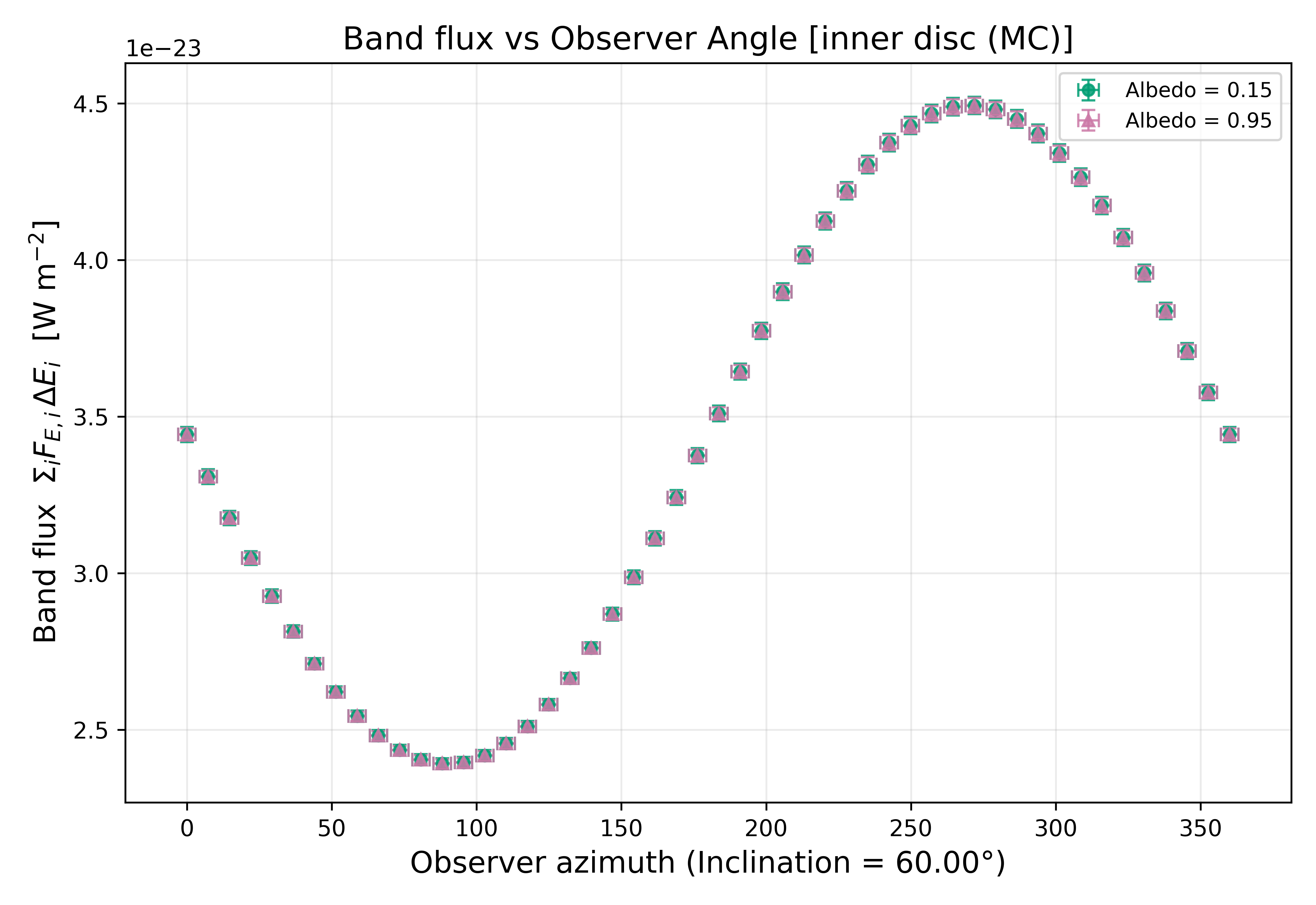}
    \caption{Flux density}
    \label{fig:Precessing_albedo_flux}
\end{subfigure}
\hfill
\begin{subfigure}{0.32\textwidth}
    \centering
    \includegraphics[width=\linewidth]{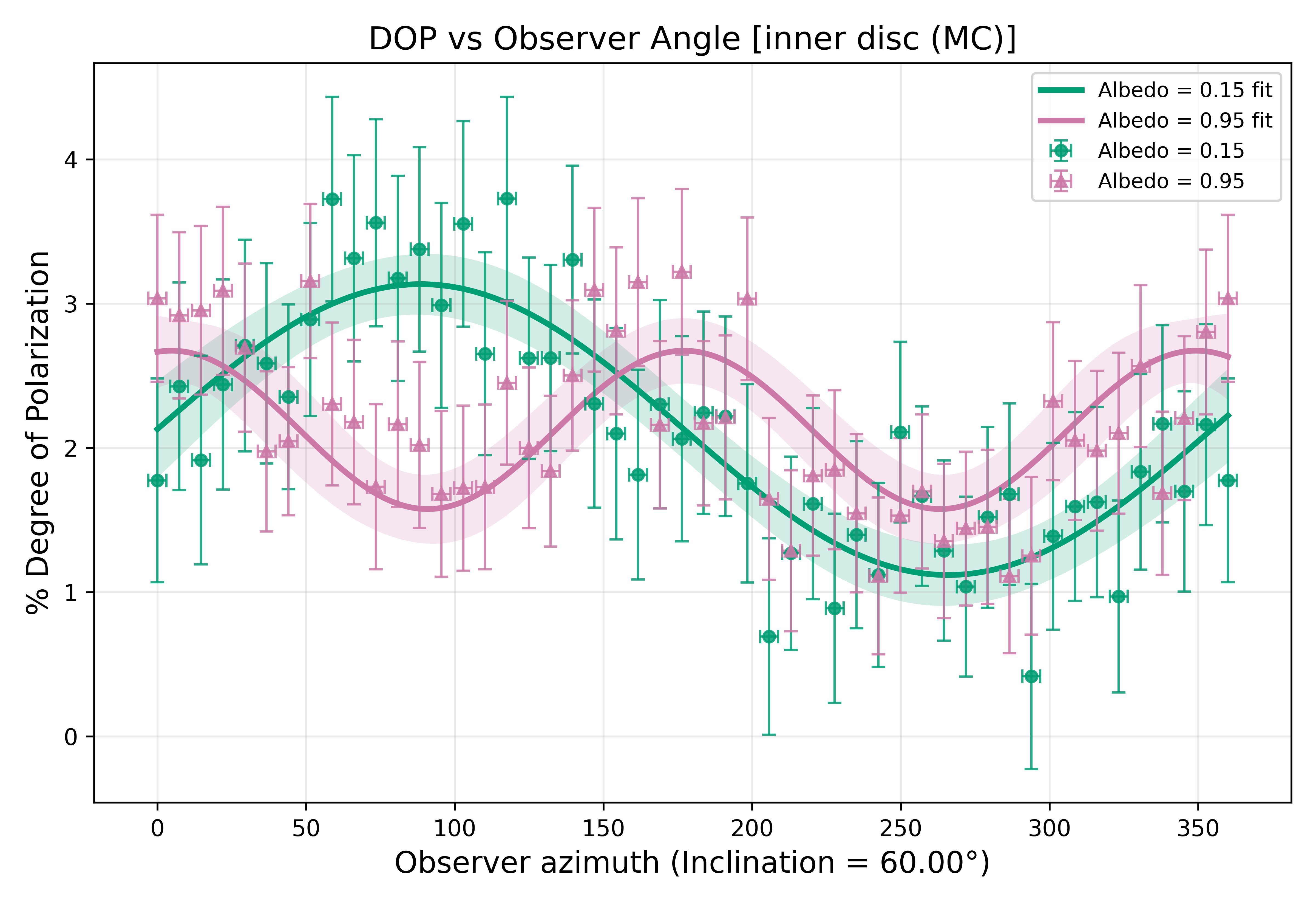}
    \caption{DOP}
    \label{fig:Precessing_albedo_DOP}
\end{subfigure}
\hfill
\begin{subfigure}{0.32\textwidth}
    \centering
    \includegraphics[width=\linewidth]{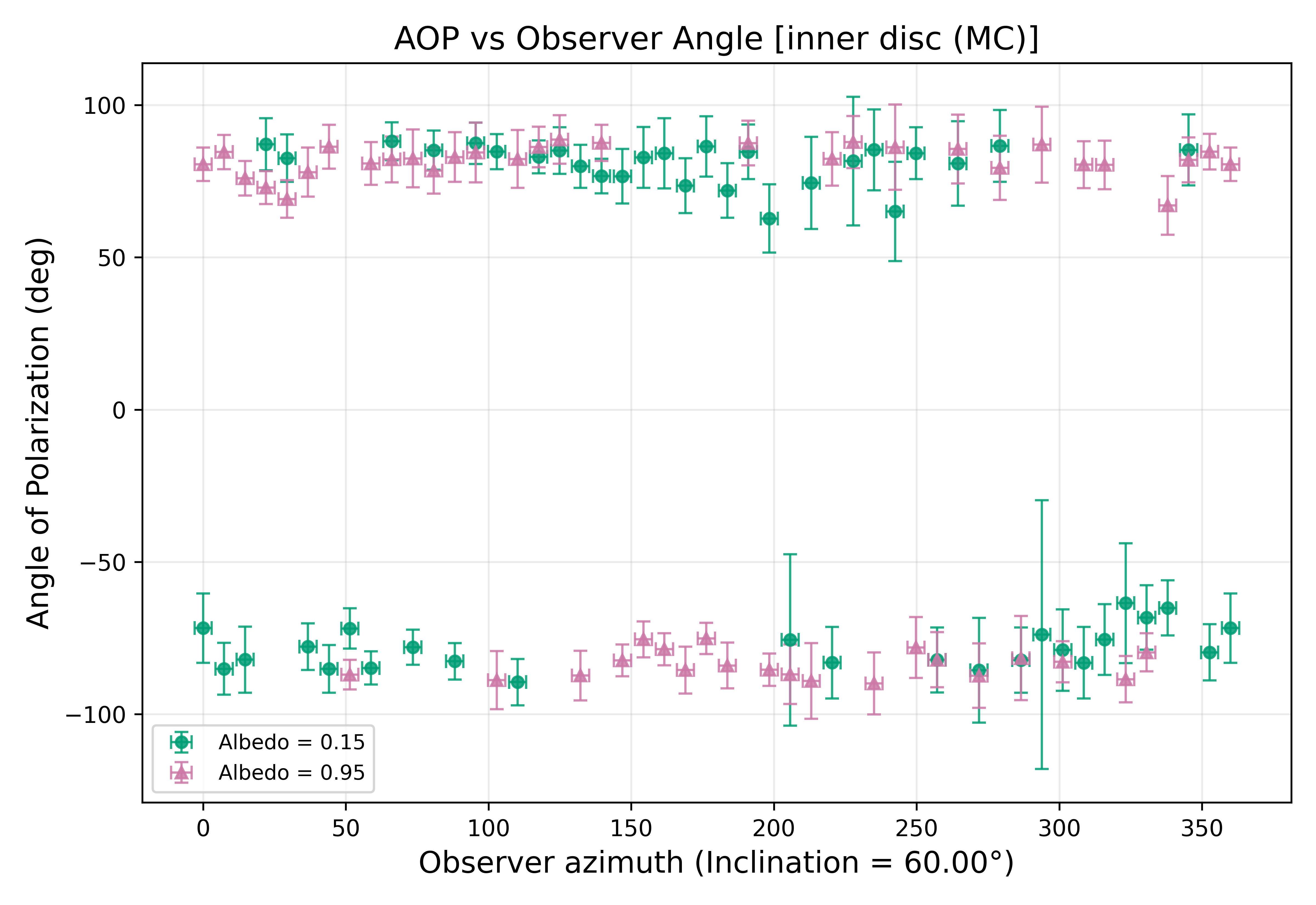}
    \caption{AOP}
    \label{fig:Precessing_albedo_DOP_AOP}
\end{subfigure}
\caption{Dependence of the phase-dependent flux density, DOP, and AOP of the inner-disc (Monte Carlo) component on the outer-disc albedo, $\rho_{\rm alb}=0.15$ and $0.95$, for the precessing inner accretion disc. The observer is at an inclination of $60^{\circ}$ with a window of $\Delta\theta\times\Delta\phi=\pm3^\circ\times\pm3^\circ$. The curves in (b) are fits of the form of Eq.~\eqref{eq:fit_prec}, with $A=1.007\pm0.0875\%$, $\omega=0.0177\pm0.0007$, $\varphi=0.003\pm0.156$, $C=2.126\pm0.059\%$\ for $\rho_{\rm alb}=0.15$ and $A=0.548\pm0.092\%$, $\omega=0.0365\pm0.002$, $\varphi=1.396\pm0.384$, $C=2.124\pm0.071\%$\ for $\rho_{\rm alb}=0.95$. The flux modulation is unchanged by the albedo, whereas the DOP modulation changes markedly and the AOP excursion is somewhat larger for the lower albedo.}
\label{fig:precessing_albedo_dependence}
\end{figure*}

We first investigate the effect of the reflection fraction by varying the albedo of the outer disc. Since the albedo determines the probability of reflection for photons incident on the outer disc, changing it modifies the contribution of reflected photons throughout the precession cycle. The resulting variations are shown in Figure~\ref{fig:precessing_albedo_dependence}. Figure~\ref{fig:Precessing_albedo_flux} shows that varying the albedo does not alter the flux-density modulation with precession phase, as was also observed for the inclination scan (Section~\ref{sec:parameter_dependence_inclination}, Figure~\ref{fig:Inc_albedo_flux}). However, the DOP in Figure~\ref{fig:Precessing_albedo_DOP} changes markedly with the reflection fraction from the outer disc: the DOP modulation is suppressed as the albedo increases, even though the flux modulation remains unchanged. The AOP modulation also differs between the two cases, with a somewhat larger deviation for the lower albedo (Figure~\ref{fig:Precessing_albedo_DOP_AOP}).

\begin{figure*}
\centering
\begin{subfigure}{0.32\textwidth}
    \centering
    \includegraphics[width=\linewidth]{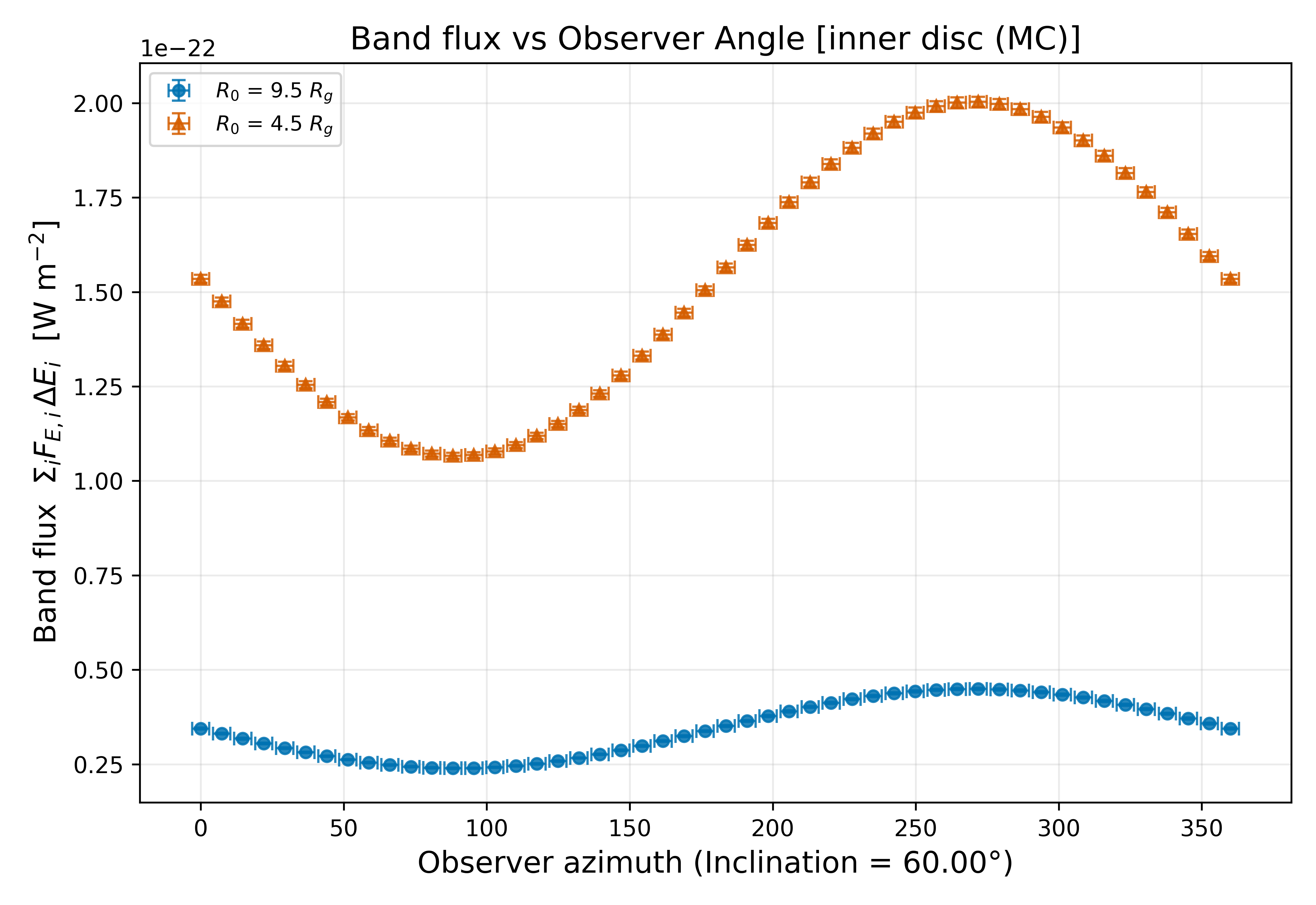}
    \caption{Flux density}
    \label{fig:Precessing_emission_radius_flux}
\end{subfigure}
\hfill
\begin{subfigure}{0.32\textwidth}
    \centering
    \includegraphics[width=\linewidth]{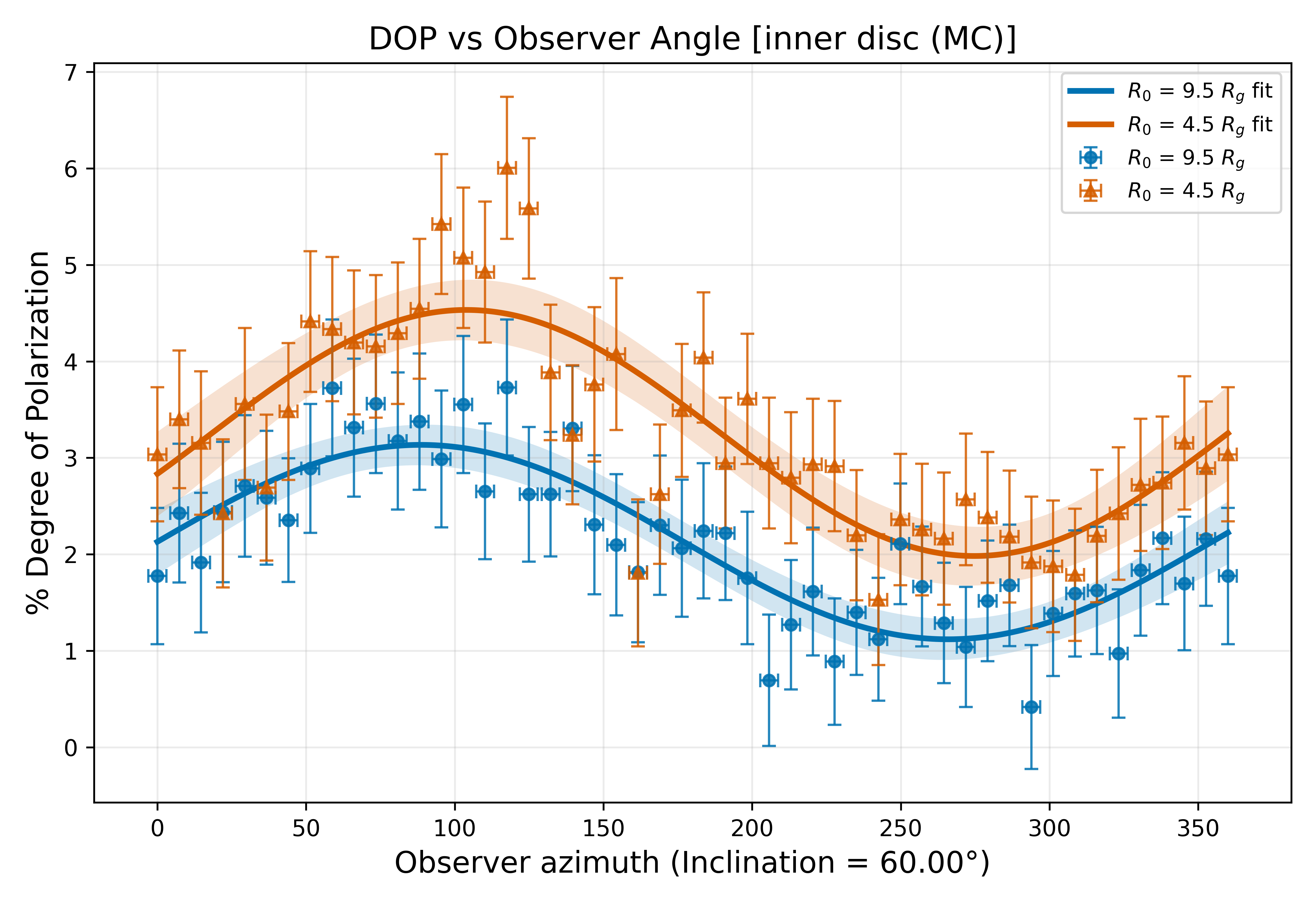}
    \caption{DOP}
    \label{fig:Precessing_emission_radius_DOP}
\end{subfigure}
\hfill
\begin{subfigure}{0.32\textwidth}
    \centering
    \includegraphics[width=\linewidth]{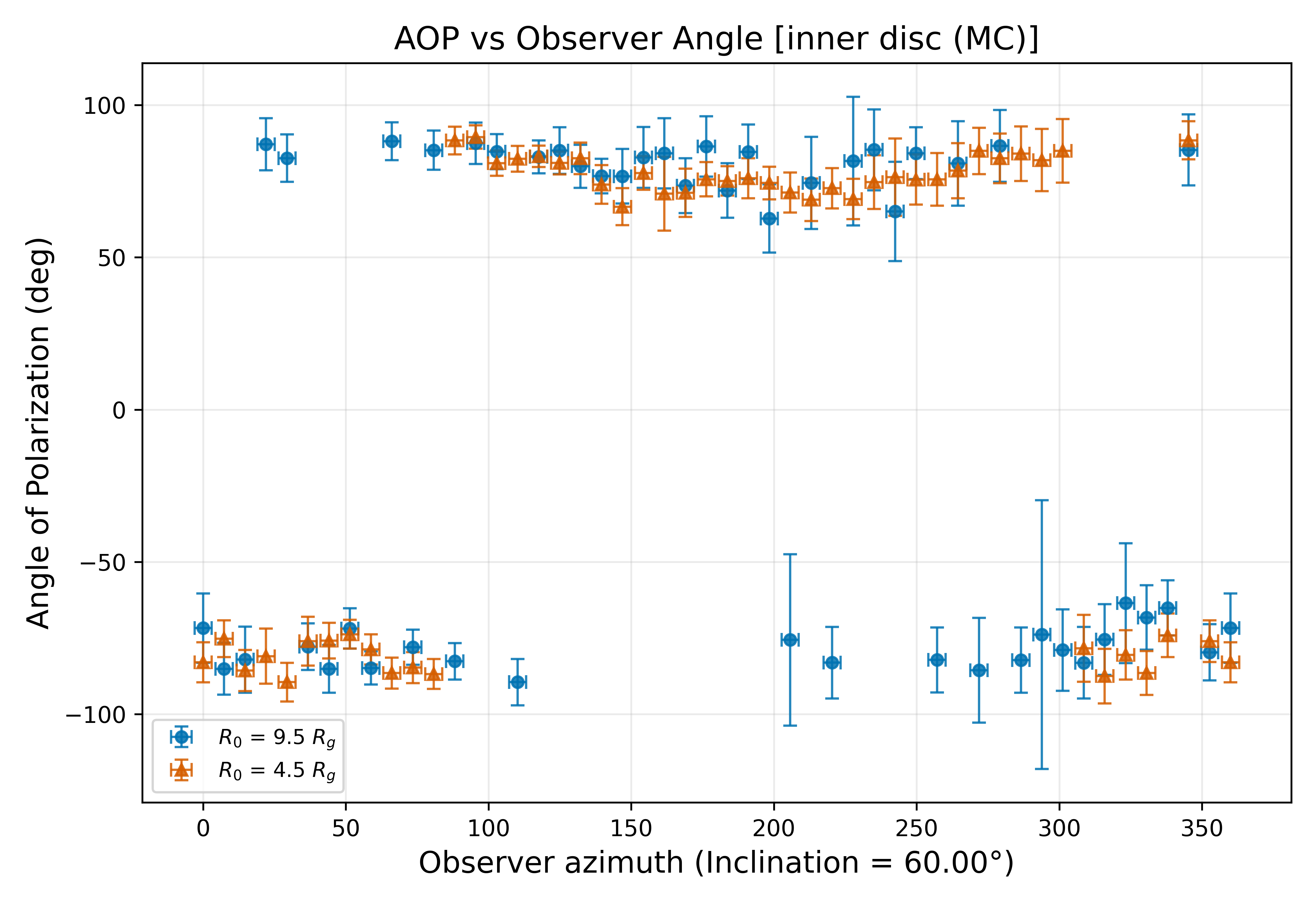}
    \caption{AOP}
    \label{fig:Precessing_emission_radius_DOP_AOP}
\end{subfigure}
\caption{Dependence of the phase-dependent flux density, DOP, and AOP of the inner-disc (Monte Carlo) component on the photon-emission radius, $R_0=4.5\,\Rg$ and $9.5\,\Rg$, for the precessing inner accretion disc. The observer is at an inclination of $60^{\circ}$ with a window of $\Delta\theta\times\Delta\phi=\pm3^\circ\times\pm3^\circ$, and the outer-disc albedo is $\rho_{\rm alb}=0.15$. The curves in (b) are fits of the form of Eq.~\eqref{eq:fit_prec}, with $A=1.275\pm0.128\%$, $\omega=0.018\pm0.0008$, $\varphi=-0.339\pm0.169$, $C=3.257\pm0.08\%$\ for $R_0=4.5\,\Rg$ and $A=1.007\pm0.087\%$, $\omega=0.018\pm0.0007$, $\varphi=0.003\pm0.156$, $C=2.126\pm0.059\%$\ for $R_0=9.5\,\Rg$. A smaller $R_0$ (stronger light bending) increases the flux modulation, the DOP modulation and the mean DOP level, and enlarges the AOP excursion from $90^{\circ}$.}
\label{fig:precessing_emission_radius_dependence}
\end{figure*}

We next examine the influence of gravitational effects by varying the photon-emission radius. The resulting variations are shown in Figure~\ref{fig:precessing_emission_radius_dependence}. The flux modulation is larger for stronger light bending (Figure~\ref{fig:Precessing_emission_radius_flux}), and so is the modulation of the DOP (Figure~\ref{fig:Precessing_emission_radius_DOP}). The overall DOP curve is also shifted upward as the light-bending effect is increased: the mean level $C$ rises from $2.13\%$ to $3.26\%$ (a factor of $\approx1.5$), while the modulation amplitude $A$ rises by a smaller factor of $\approx1.3$ (from $1.01\%$ to $1.28\%$), as can be verified from the fits in the caption of Figure~\ref{fig:precessing_emission_radius_dependence}. Figure~\ref{fig:Precessing_emission_radius_DOP_AOP} shows that stronger light bending corresponds to a larger modulation of the AOP away from its standard value of $90^{\circ}$ for the corresponding observer.

\begin{figure*}
\centering
\begin{subfigure}{0.32\textwidth}
    \centering
    \includegraphics[width=\linewidth]{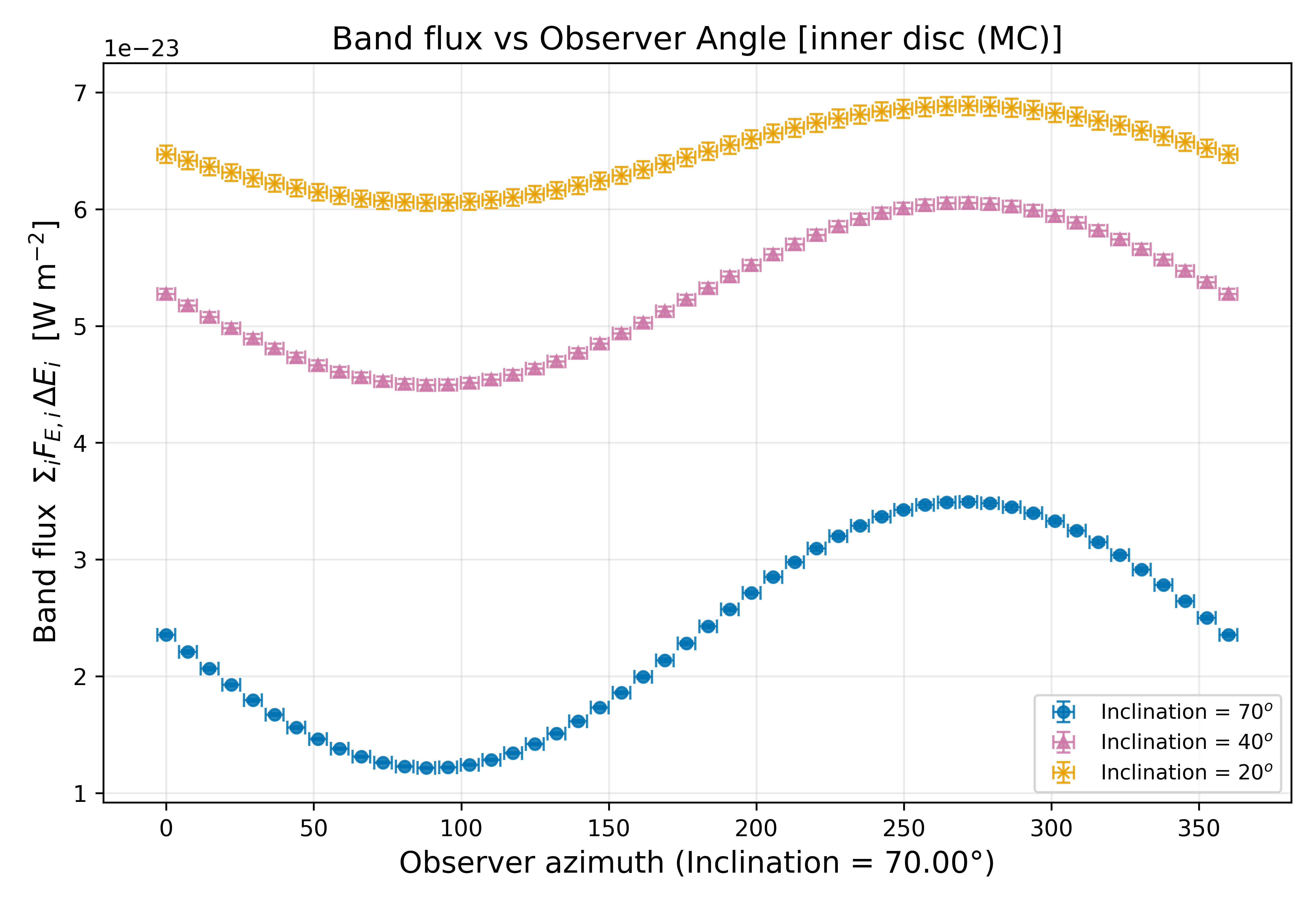}
    \caption{Flux density}
    \label{fig:Precessing_inclination_flux}
\end{subfigure}
\hfill
\begin{subfigure}{0.32\textwidth}
    \centering
    \includegraphics[width=\linewidth]{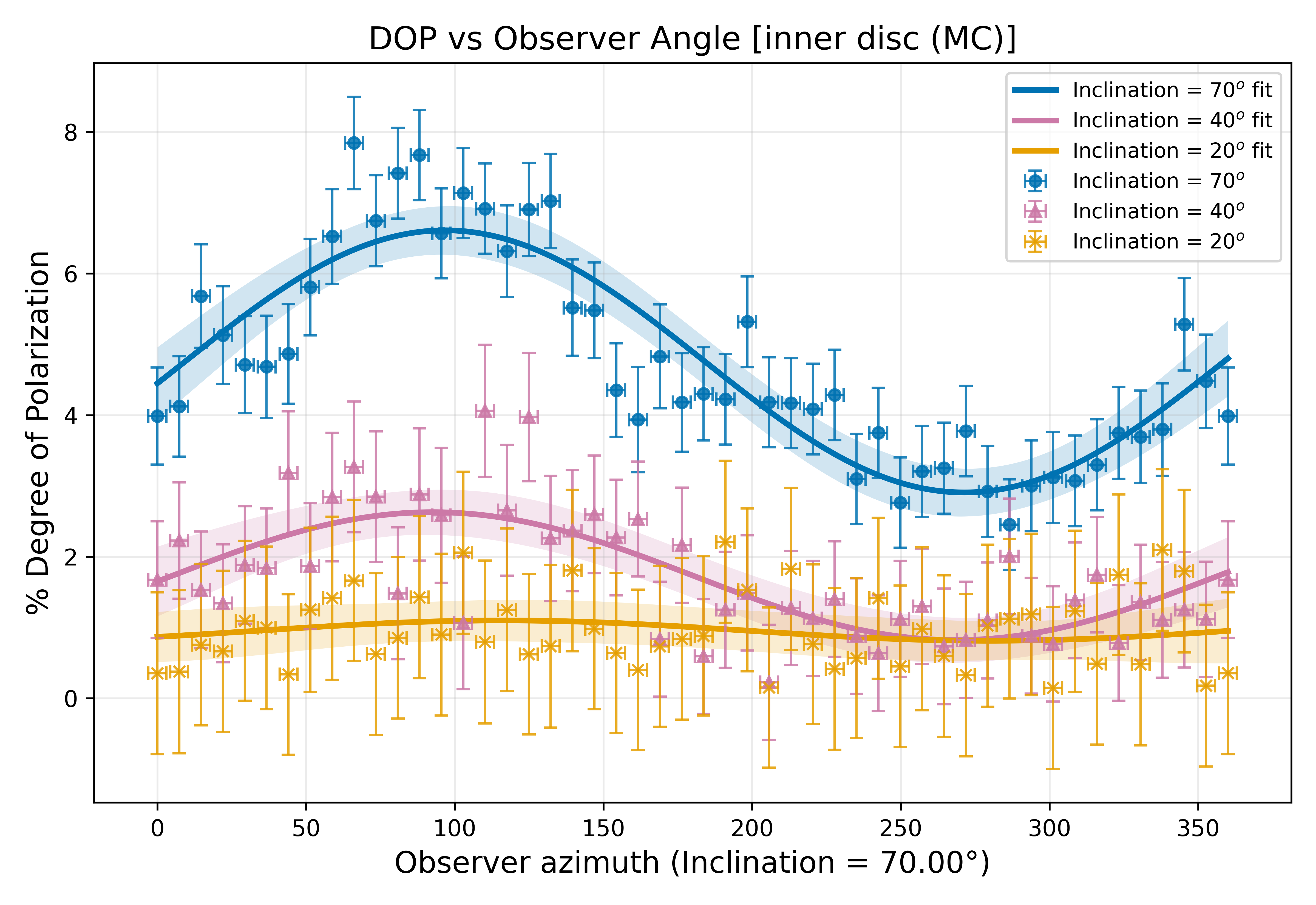}
    \caption{DOP}
    \label{fig:Precessing_inclination_DOP}
\end{subfigure}
\hfill
\begin{subfigure}{0.32\textwidth}
    \centering
    \includegraphics[width=\linewidth]{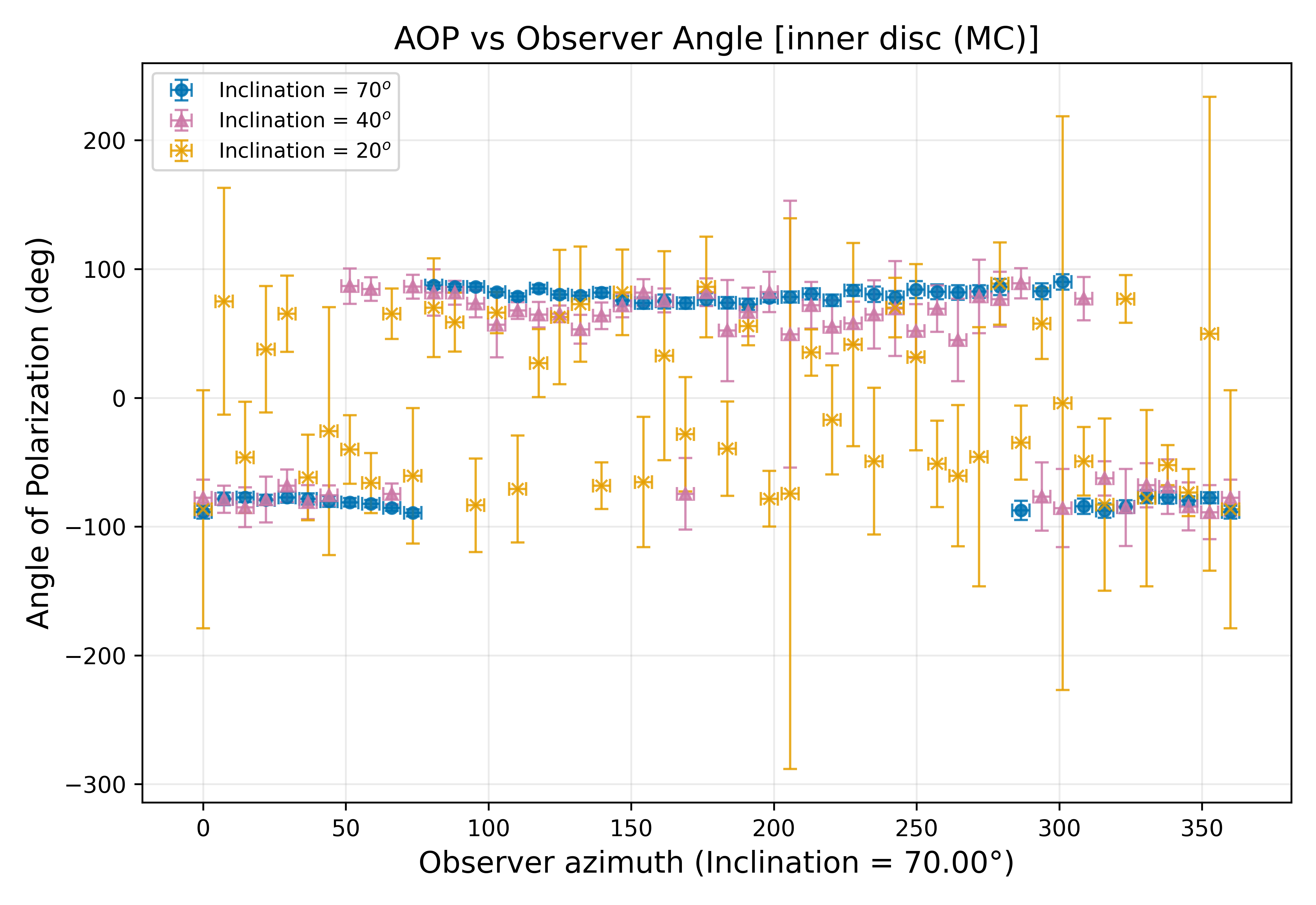}
    \caption{AOP}
    \label{fig:Precessing_inclination_DOP_AOP}
\end{subfigure}
\caption{Dependence of the phase-dependent flux density, DOP, and AOP of the inner-disc (Monte Carlo) component on observer inclination, $i=20^{\circ}$, $40^{\circ}$, and $70^{\circ}$, for the precessing inner accretion disc. Photons are emitted from $R_0=4.5\,\Rg$, the outer-disc albedo is $\rho_{\rm alb}=0.15$, and the observer window is $\Delta\theta\times\Delta\phi=\pm3^\circ\times\pm3^\circ$. The curves in (b) are fits of the form of Eq.~\eqref{eq:fit_prec}, with $A=0.143\pm0.117\%$, $\omega=0.019\pm0.006$, $\varphi=-0.66\pm1.36$, $C=0.956\pm0.081\%$\ for $i=20^{\circ}$, $A=0.903\pm0.132\%$, $\omega=0.018\pm0.0012$, $\varphi=-0.082\pm0.256$, $C=1.725\pm0.089\%$\ for $i=40^{\circ}$, and $A=1.851\pm0.141\%$, $\omega=0.018\pm0.0006$, $\varphi=-0.169\pm0.131$, $C=4.758\pm0.095\%$\ for $i=70^{\circ}$. The flux modulation increases with inclination, the DOP modulation is almost absent at $20^{\circ}$ and well defined at $40^{\circ}$ and $70^{\circ}$, and the AOP spread decreases with inclination.}
\label{fig:precessing_inclination_dependence}
\end{figure*}

Lastly, we investigate the effect of the observer inclination on the polarization modulation produced by the precession of the inner accretion disc. Different observer inclinations correspond to different viewing geometries of the inner disc, and therefore modify the amplitude of the polarization signatures over a precession cycle. The resulting phase-dependent flux density, DOP, and AOP for different observer inclinations are shown in Figure~\ref{fig:precessing_inclination_dependence}. Figure~\ref{fig:Precessing_inclination_flux} shows that the flux-density modulation increases with increasing observer inclination, although the overall flux level decreases. Figure~\ref{fig:Precessing_inclination_DOP} shows that a clear, sinusoidal DOP modulation is observed for highly inclined observers, but not for low-inclination observers. For an inclination of $20^{\circ}$ the modulation is almost invisible and the curve is close to a straight line, in contrast to the well-defined modulations at inclinations of $40^{\circ}$ and $70^{\circ}$. The spread in the AOP is larger for observers at lower inclination and decreases gradually for more highly inclined observers (Figure~\ref{fig:Precessing_inclination_DOP_AOP}).

These comparisons allow the polarization modulation produced by inner-disc precession to be examined under different reflection fractions, different strengths of gravitational effects, and different observer viewing geometries. The corresponding flux variations are considered alongside the DOP and AOP to characterize the combined changes in the intensity and polarization of the observed radiation throughout the precession cycle.
\subsection{Precession-induced effects on the DOP and AOP of the whole system}
\label{sec:whole_system}

Finally, we extend the analysis from the inner accretion disc to the complete disc system. We now consider the composite radiation obtained by combining the contributions from the inner and outer disc. The resulting polarization signatures therefore contain the combined contribution of the radiation emerging from the inner disc together with the additional analytic emission associated with the outer disc.

\begin{figure*}
\centering
\begin{subfigure}{0.48\textwidth}
    \centering
    \includegraphics[width=\linewidth]{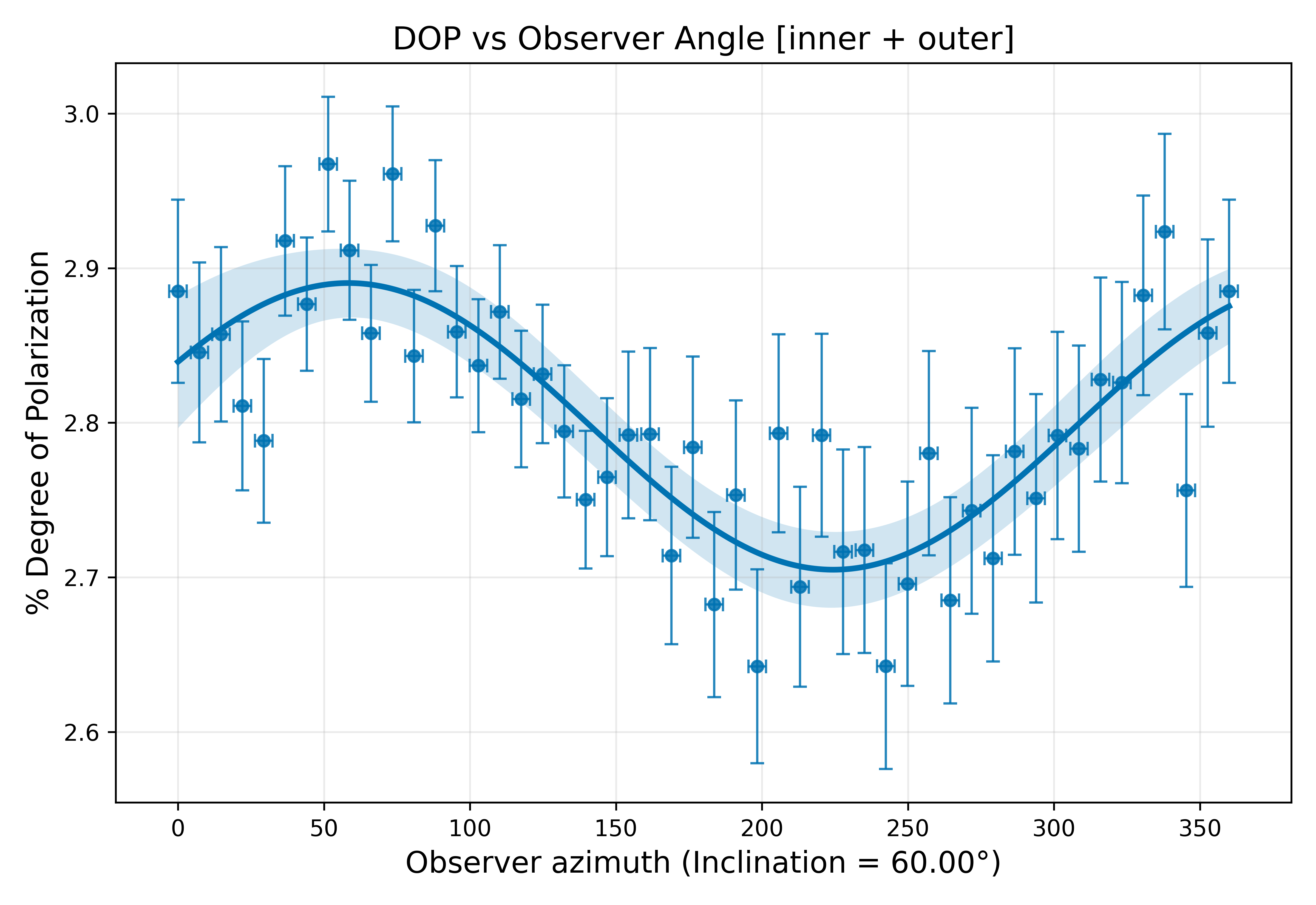}
    \caption{DOP of the whole system}
    \label{fig:whole_disk_DOP}
\end{subfigure}
\hfill
\begin{subfigure}{0.48\textwidth}
    \centering
    \includegraphics[width=\linewidth]{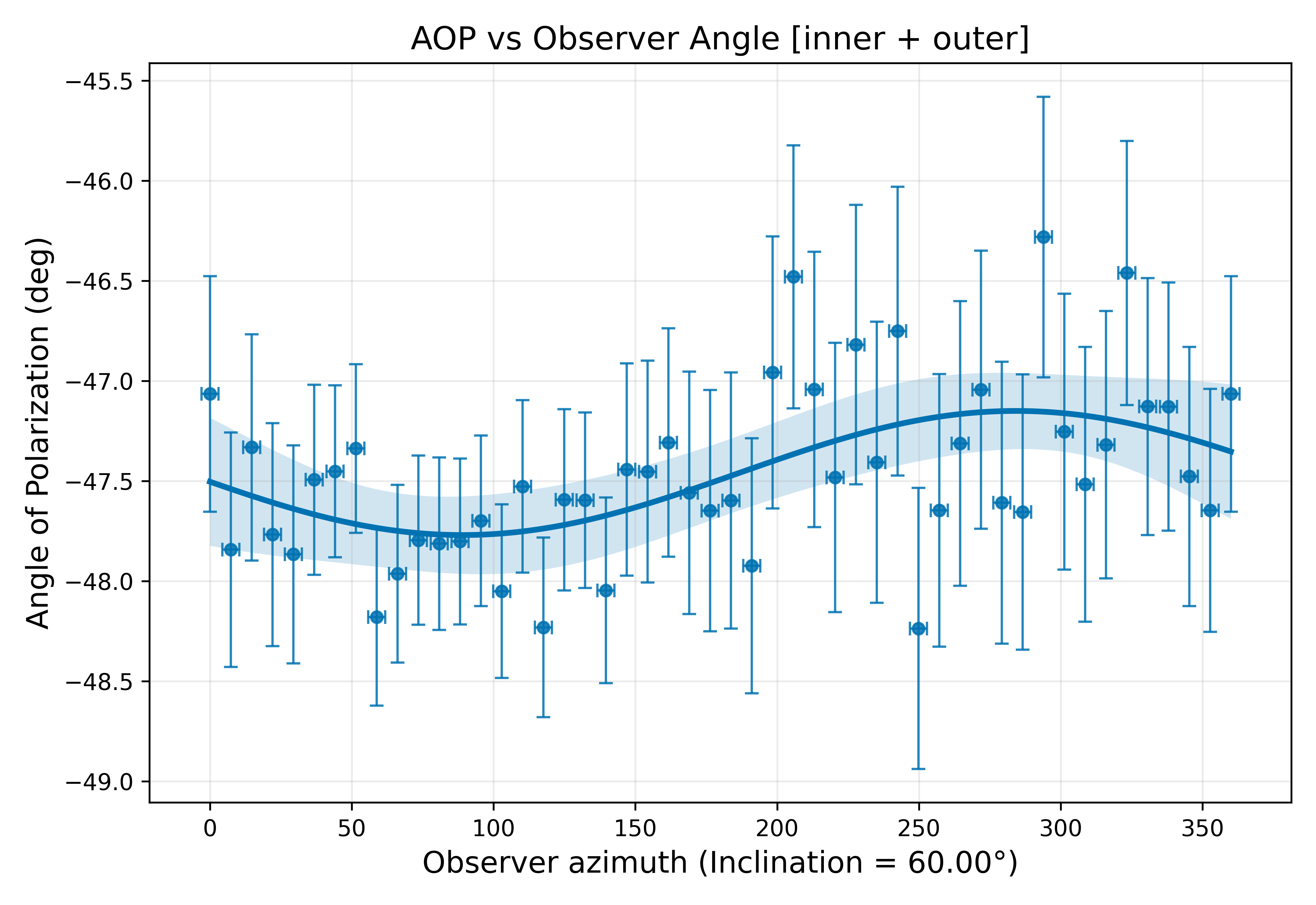}
    \caption{AOP of the whole system}
    \label{fig:whole_disk_AOP}
\end{subfigure}
\caption{Variation of the DOP and AOP of the complete disc model, including the contributions from photons originating from both the inner and outer disc, with precession phase. The observer is at an inclination of $60^{\circ}$ with respect to the BH spin axis, starting from an azimuth of $0^{\circ}$, with a window of $\Delta\theta\times\Delta\phi=\pm3^\circ\times\pm3^\circ$, $R_0=9.5\,\Rg$ and $\rho_{\rm alb}=0.15$. The curve in (a) is a fit of the form of Eq.~\eqref{eq:fit_prec}, with $A=0.0926\pm0.009\%$, $\omega=0.0188\pm0.001$, $\varphi=0.486\pm0.222$, $C=2.798\pm0.007\%$. The outer-disc emission is steady and dilutes the phase-dependent inner-disc modulation.}
\label{fig:whole_disk_polarization}
\end{figure*}

Figure~\ref{fig:whole_disk_polarization} shows the resulting DOP and AOP for the complete disc model. The observer is initially positioned at an inclination of $60^{\circ}$ with respect to the BH spin axis and an azimuth of $0^{\circ}$ with respect to the $X$ axis of the system. We observe that the DOP and AOP vary only within a narrow band (centred on approximately $2.8\%$ and $-47.5^{\circ}$; and having an amplitude of 0.09\% and 0.3$^\circ$ respectively), which would in practice be undetectable by any of the currently operating polarimetric satellites. This is understandable, since the fluctuations originating from the inner disc are superimposed on the effectively constant DOP and AOP contributed by the outer disc. The extent to which the outer-disc contribution dilutes that of the inner disc depends on the relative photon-weighted contributions of the two regions, which in turn depend on the size and location of the inner disc's photon-emitting region relative to the outer disc. The physical parameters can therefore be tuned to identify conditions under which the overall DOP and AOP modulation of the system yields a detectable contribution from both the inner and outer discs.

\section{Conclusions}
\label{sec:conclusions}

We have extended the GRMC radiative transfer model of Paper~I to self-consistently track the full photon polarization state, combining the precessing, Lense--Thirring-driven inner accretion flow, the hot spherical Comptonizing corona, and the misaligned, Bardeen--Petterson-truncated outer disc within a single Kerr-spacetime calculation. By adding a polarization 4-vector to every photon, parallel-transported via the Walker--Penrose formalism and scattered through a polarized Compton and disc-reflection treatment, the model links the energy- and phase-resolved DOP and AOP to the same energy spectra, path-length distributions, and spectral lags computed in Paper~I, providing a unified diagnostic linking timing, spectroscopy, and polarimetry.

Gravitational light bending raises the DOP: with scattering switched off, introducing the black hole increases the polarization amplitude relative to flat spacetime, because capture by the horizon removes the strongly bent, least polarized photon paths. Coronal scattering has the opposite effect, lowering the DOP as the optical depth increases, and reflection from the outer disc also lowers the DOP. For a precessing inner flow, stronger light bending increases the polarization modulation, while photons reflected from the misaligned outer disc suppress it, and the steady polarization of the outer disc further dilutes the modulation of the composite system. These dependencies provide a general framework for interpreting anomalous polarimetric behaviour, since the relative strength of the competing effects determines whether the emergent radiation is weakly or strongly polarized and modulated. As illustrations, a scattering-rich environment with reflection and outer-disc dilution gives a low mean DOP and a nearly null modulation, as in Swift~J1727.8$-$1613, whereas little secondary scattering combined with strong light bending gives a high DOP, as in IGR\,J17091$-$3624. The observer's inclination also plays a role, since it controls both the mean DOP and the amplitude of the precession-driven modulation.

The energy-resolved polarization signal carries a clear imprint of the two-zone geometry. At low energies, where the outer disc dominates the observed flux, the DOP approaches the value expected from Chandrasekhar's semi-infinite electron-scattering atmosphere for the corresponding disc inclination ($\approx\!6\%$ for our fiducial observer), and the AOP sits at the canonical $90^{\circ}$ set by the disc normal. As energy increases and the inner-disc, corona-processed photons come to dominate, the DOP drops to $\approx\!0.5\%$, well below the value a naive, scattering-free estimate would predict for the inner disc's own tilt, and the AOP rotates towards the projection of the spin axis on the sky. This deviation is direct evidence that coronal scattering and gravitational light bending act together to depolarize and rotate the emergent radiation relative to simple geometric expectations.

Isolating the two mechanisms shows that they act in opposite senses on the mean DOP. The inclination dependence of the DOP of the inner-disc component follows the Chandrasekhar form, peaking for an edge-on view of the tilted inner disc. In flat spacetime, increasing the coronal optical depth increases the number of Compton scatterings per escaping photon and systematically lowers the DOP, as expected from the repeated reorientation of the polarization vector: the fitted amplitude falls from $11.0\pm0.6\%$ with scattering switched off to $6.9\pm0.8\%$ and $2.3\pm0.7\%$ for electron-density normalizations of $1\times10^{22}$ and $3\times10^{22}\,\mathrm{m^{-3}}$, respectively. In contrast, with scattering switched off, introducing the black hole \emph{raises} the amplitude from $11.0\pm0.6\%$ (consistent with the Chandrasekhar maximum of $11.71\%$), to $14.3\pm0.7\%$, at an unchanged flux. In all cases the AOP stays at its expected value except at extreme inclinations, where its scatter grows.

Systematically varying the disc albedo and the radius of photon injection isolates the two competing physical mechanisms that control the polarization. Reflection from the outer disc, being Compton scattering, depolarizes the radiation: raising the albedo from 0.15 to 0.95 lowers the fitted amplitude from $9.4\pm0.6\%$ to $6.3\pm0.6\%$ and makes the DOP profile asymmetric about its peak, without changing the flux. Conversely, moving the photon-emission region inwards, from $9.5\,\Rg$ to $4.5\,\Rg$, and thus increasing the strength of light bending experienced by the escaping photons, raises the amplitude from $8.8\pm0.5\%$ to $11.0\pm0.5\%$ even in the presence of coronal scattering. The same trends carry over to the precessing flow. Stronger light bending increases the flux modulation, the DOP modulation, and the AOP excursion, and shifts the DOP curve upwards, demonstrating that strong-field light bending amplifies the polarimetric signature of inner-disc precession, whereas the reflected photons from the misaligned outer disc suppress it without altering the flux modulation. The net observable modulation is set by the competition between these two effects, modulated further by the observer's inclination: at $40^{\circ}$ and $70^{\circ}$ the DOP shows a clean, quasi-sinusoidal modulation with precession phase, while at $20^{\circ}$ the modulation is almost absent and the AOP spread is larger.

When the full system is considered, combining the phase-dependent inner-disc signal with the steady, comparatively bright emission of the misaligned outer disc, the two contributions do not add constructively. The outer disc's constant polarization state dilutes the precession-driven modulation of the inner disc, narrowing the net observable range of DOP and AOP swept out over a precession cycle. As an illustration, this dilution offers a natural, physically motivated explanation for the low modulation reported for Swift~J1727.8$-$1613: relativistic precession models that consider the inner flow in isolation predict a polarimetric modulation that current IXPE observations do not detect, whereas a composite model that accounts for coronal scattering, depolarization by reflection, and dilution by a misaligned outer disc naturally produces a much smaller net signal, more consistent with the reported near-null amplitudes. Coronal scattering and reflection also lower the mean DOP, which could keep it near the observed $\sim\!4\%$ even at high inclination, where the simple precession picture would otherwise push it to $\gtrsim7\%$ \citep{Zhao2024}.

The same ingredients, differently weighted, can also account for the opposite situation, in which an unusually high DOP is observed, as in IGR\,J17091$-$3624 \citep{Ewing2025}. Our results suggest that this can happen when there is no significant secondary scatterer, such as the spherical corona considered here, around the inner flow, so that the depolarization caused by multiple scatterings (Section~\ref{sec:corona_density_dependence}) is absent. In that case the DOP is instead amplified by the presence of the black hole itself, which removes the strongly bent, least polarized photon paths from the observed ensemble (Section~\ref{sec:light_bending_only}), and this is further favoured by a low reflection fraction, a compact emitting region with strong light bending, and a near edge-on view. This is a qualitative argument. \citet{Ewing2025} additionally invoke a favourable coronal geometry, a high inclination, and a possible outflow or optically thick wind, none of which is included in our spherical-corona model, and a quantitative comparison would require fitting the model to the data. The same configuration also gives the largest precession-induced modulation, as it combines strong light bending, low reflection, and no outer-disc dilution. We therefore tentatively predict that, if the $\sim\!0.2$\,Hz type-C QPO of IGR\,J17091$-$3624 is driven by precession, its polarization should be modulated more strongly with QPO phase than that of Swift~J1727.8$-$1613.

These results have direct implications for the interpretation of existing and forthcoming X-ray polarimetric data. A non-detection of QPO-phase-resolved polarimetric modulation cannot, on its own, be taken as evidence against a precessing inner flow: the suppression mechanisms identified here mean that the RPM remains viable even where its naive prediction is excluded. Conversely, where a modulation \emph{is} detected, or an unusually high mean DOP is measured, its amplitude, energy dependence, and phase relative to the flux and spectral variability of Paper~I can be used, within this combined framework, to constrain the relative strengths of coronal scattering, gravitational light bending, and outer-disc reflection, and hence the disc-warp geometry and observer inclination of the system. This provides a quantitative basis for future joint timing, spectral, and polarimetric campaigns with IXPE, and for the enhanced sensitivity anticipated from XPoSat and the upcoming eXTP mission.

The precession of the inner disc in this work was modelled by exploiting a kinematic equivalence: rather than rotating the inner disc about the spin axis at every phase and repeating the Monte Carlo simulation, we sweep the observer azimuth around the spin axis for a single photon dataset. This is a valid, though approximate, way of treating the problem, and it is adequate to first order. The Kerr spacetime and the spherical corona are axisymmetric about the spin axis, so rotating the inner disc by an angle $\varphi$ is equivalent to rotating the observer by $-\varphi$, and the phase dependence of the inner-disc emission, including its Comptonization and light bending, is recovered without further approximation. This scheme is what makes the extensive parameter study presented here practical: a single simulation provides every precession phase, reducing the computation time by roughly the number of phases sampled, and it allowed us to map the dependence of the polarization modulation on the coronal density, albedo, emission radius, and inclination, with no loss of accuracy at the first-order level considered here. The one ingredient that the scheme treats approximately is the outer disc. Because it is held fixed, the direction from which the outer disc is illuminated by the inner flow is the same at every phase, whereas in a real precessing system it changes over the cycle. We expect this to be a small correction to the main results, which are heavily driven by the light bending, coronal scattering, and the steady outer-disc emission; although we have not quantified it in this paper.

The model itself is not limited by this approximation, and can be run seperately for multiple phases of the inner disc, with the disc physically rotated and the simulation repeated at each phase, to give a fully accurate picture, in particular for a precise comparison with phase-resolved observations. Several other simplifications in the current model also point to clear directions for future work, most shared with Paper~I: a continuously warped disc profile rather than two rigid zones, a GRMHD-informed corona rather than a static spherical one, and a full ionized-reflection treatment of the outer disc rather than a Thomson-scattering albedo prescription would all sharpen the predictions presented here. The corona is currently restricted to a spherically symmetric geometry, and we plan of extending the framework to slab, lamp-post, and outflowing configurations, to test how sensitively the predicted polarization and timing signatures depend on the assumed coronal geometry; which would enable a more direct comparison with the corona models commonly invoked in the X-ray spectral-timing literature, and would allow a quantitative test of the geometries invoked to explain the high DOP of IGR\,J17091$-$3624. Although the corona is formulated with a radially varying electron density profile, the current photon-transport implementation approximates this density as constant along each scattering path. A possible refinement would incorporate the local variation of electron density along the photon trajectory directly into the free-path sampling and scattering probability, providing a more physically consistent treatment of Compton scattering within a radially stratified corona.

Several further extensions specific to the polarization signal may also be considered. The current treatment of outer-disc reflection does not include line emission. Incorporating the fluorescent iron K$\alpha$ (and associated) line complex into the emergent energy spectrum would allow the model to be compared directly against the broad iron-line diagnostics widely used to probe disc geometry and black hole spin. Closely related, the polarization signatures of this reflected spectral component could be computed in their own right, since the DOP and AOP of the reflected continuum and line emission are expected to carry additional, energy-resolved information about the outer-disc geometry beyond what is captured by the continuum polarization alone.

\section*{Acknowledgements}

MC and DB thank Ashoka University and Axis Bank for funding this research via the Axis Bank PhD grant. The Monte Carlo simulations presented in this paper, as well as in Paper~I, were carried out using three computing facilities at Ashoka University: the Gsharp cluster (courtesy of the Physics Department), Tao (courtesy of Prof. Garima Mishra, with access facilitated by Souradeep Sengupta and Bidisha Mukherjee), and one computational facility of Prof. Suratna Das (with access facilitated by her PhD student Debadyuti Ghosh). We also acknowledge the use of AI tools (Claude, ChatGPT, and Gemini) in the preparation of this manuscript, primarily for language and text-level editing.

\section*{Code Availability}

The simulation code underlying this article will be shared on
reasonable request to the corresponding author.

\bsp
\label{lastpage}
\end{document}